\documentclass[preprint,aps,showpacs,preprintnumbers,amsmath,amssymb,superscriptaddress]{revtex4-1}

\usepackage{graphicx}
\usepackage{dcolumn}
\usepackage{multirow}
\usepackage{placeins}
\usepackage{xcolor}
\usepackage[title]{appendix}
\usepackage{soul}

\begin{document}


\title{Revisiting magnetic exchange interactions in transition metal doped Bi$_2$Se$_3$ using DFT+MFT}
\author{Sagar Sarkar}
\affiliation{Asia Pacific Center for Theoretical Physics, Pohang, 37673, Korea}
\affiliation{Department of Physics and Astronomy, Uppsala University, Uppsala, 751 20, Sweden}
\author{Shivalika Sharma}
\affiliation{Asia Pacific Center for Theoretical Physics, Pohang, 37673, Korea}
\affiliation{Institute of Physics, Nicolaus Copernicus University, 87-100 Toruń, Poland}
\author{Igor Di Marco}
\affiliation{Asia Pacific Center for Theoretical Physics, Pohang, 37673, Korea}
\affiliation{Institute of Physics, Nicolaus Copernicus University, 87-100 Toruń, Poland}
\date{January 2024}

\begin{abstract}
    Topological insulators doped with magnetic impurities has become a promising candidate for Quantum  Anomalous Hall Effect (QAHE) in the dilute doping limit. The crucial factor in realizing the QAHE in these systems is the spontaneous Ferromagnetic (FM) ordering between the doped magnetic atoms. Hence, understanding the magnetic exchange interaction between the magnetic atoms becomes essential.  In this work, we use the Density functional theory (DFT) and Magnetic force theorem (MFT) to calculate the magnetic exchange interaction between magnetic impurities (V, Cr, Mn, Fe) in the host Bi$_2$Se$_3$. Through an orbital decomposition of the calculated exchange, we can identify the nature and origin of the exchange mechanism that depends on the type of magnetic atoms, doping concentration, host material etc. Our results show that Cr doping results in an insulating state, a prerequisite for the QAHE,  that remains robust against doping concentration and local correlation. In this case, the short-ranged superexchange and long-ranged exchange via the $p$-orbitals of the host results in an FM order. For other doped systems (V, Mn and Fe doped), their electronic configuration and local octahedral environment open the possibility of finite carrier density at the Fermi energy. Depending on the type of this carrier (electron/hole) and their localized/delocalized nature, a short-ranged double exchange / long-ranged RKKY mechanism could occur between the magnetic atoms.  
\end{abstract}

\maketitle

\section{Introduction}

The presence of quantized Landau levels in a 2D electron gas under a strong magnetic field leads to a topological state of matter with zero longitudinal resistance in association with a
quantized Hall conductance. This is called the quantum Hall effect (QHE) first shown experimentally by von Klitzing~\cite{klitzing_PhysRevLett.45.494_1980}. This state forms a bulk bandgap for this 2D system,  but the boundaries or edges exhibit 1D gapless chiral edge states (CES). These CES are protected from back-scattering (from one edge to another) allowing for a unidirectional, dissipationless flow of current along the edges of the sample~\cite{liu_PhysRevLett.101.146802_2008,wu_PhysRevLett.113.136403_2014}. 
These features make such a system promising for electronic devices. Interestingly, such a state does not necessarily require the presence of an external magnetic field. Instead, they rely on the presence of protected CES with broken time-reversal symmetry(TRS)\cite{liu_PhysRevLett.101.146802_2008,wu_PhysRevLett.113.136403_2014}. Theoretical predictions~\cite{haldane_PhysRevLett.61.2015_1988,liu_PhysRevLett.101.146802_2008,rui_science.329.5987_2010} suggest that such a state can be realized in certain 2D magnetic insulating materials due to their internal/intrinsic magnetic field, leading to Quantum anomalous Hall effect (QAHE)~\cite{zhang_PhysRevB.93.235315_2016}. Hence, to achieve QAHE, two essential conditions have been identified ~\cite{rui_science.329.5987_2010}(i), a 2D insulating state with FM order to break TRS and (ii) a band inversion transition resulting from spin-orbit coupling (SOC) to have CES.
These conditions make topological insulators (TI) a good candidate for QAHE. TI has a bulk bandgap resulting from SOC along with topologically protected gap-less surface states with TRS~\cite{zhang_natphy_5.6_2009}. Hence, establishing an FM order in a suitable TI (for example Bi$_2$Se$_3$) will break the TRS and then by tuning the Fermi level of the sample around the magnetically opened energy gap will naturally lead to QAHE~\cite{cui_science.340.167_2013}. This has been experimentally demonstrated in Cr/V doped films of Bi$_2$Se$_3$ family of compounds (Bi, Sb)$_2$Te$_3$~\cite{cui_science.340.167_2013,kou_PhysRevLett.113.137201_2014,checkelsky_natphy.10.731_2014,bestwick_PhysRevLett.114.187201_2015,chang_natphy.14.473_2015}. Similar to QHE, here also dissipationless current flows via the 1D CES but in absence of an external magnetic field. This makes QAHE systems a potential candidate for electronic, spintronic devices~\cite{liu_PhysRevLett.101.146802_2008,zhang_PhysRevB.93.235315_2016}, and quantum computing~\cite{chen_PhysRevB.97.104504_2018}. For a real device application of QAHE, the temperature stability of this magnetic state with topological properties is essential. 
Until now, QAHE has been observed at low temperatures\cite{cui_science.340.167_2013,chang_natphy.14.473_2015,bestwick_PhysRevLett.114.187201_2015,YFeng_PRL_115_2015} generally in the range of 15-30 K\cite{CChang_RevModPhys_95_2023}. Achieving full quantization requires even lower temperature, below 300mk, as reported in previous studies \cite{OYunbo_AdvMater_30_2017,MMogi_APL_107_2015}, which is significantly lower compared to the ferromagnetic ordering temperature (~100K) of these systems\cite{ZZhou_APL_87_2005,ZZhou_PRB_74_2006,chang_natphy.14.473_2015}. Hence, it is crucial to understand the microscopic origin of magnetic exchange between the dopants, its dependence on the dopant type, and doping concentration, and their effect on the topological band structure or electronic structure of the host material.
Depending on the dopant type, doping site, and doping concentration various types of exchange mechanisms can occur in these magnetically doped topological insulators. As a doped system, they are very similar to dilute magnetic semiconductors (DMS),  where free carrier-driven RKKY is considered to be the dominant mechanism for long-ranged magnetic exchange\cite{AWerpachowska_arxiv.1111.2011}, for example in Mn-doped CdSe/Te\cite{TDietl_PRB.55_1997}. However, for doped topological insulators with insulating properties, such as Cr and Fe doped Bi$_2$Se$_3$\cite{YRui_science.329_2010} the van Vleck mechanism\cite{VVlek_book_2015} was suggested that do not require the presence of free carriers. This mechanism depends on the strength of spin-orbit coupling (SOC) and an enhanced spin susceptibility  results from the inverted band structure of the TIs\cite{YRui_science.329_2010,CCui_AdvMat.25_2013}. The signature of van Vleck type exchange was also experimentally demonstrated later for these types of systems\cite{CCui_AdvMat.25_2013,LMingda_PRL.114_2015}. However, later theoretical studies on magnetically doped TIs with finite carrier density at the Fermi level suggested the importance of the long-ranged RKKY mechanism in these systems\cite{verginory_PhysRevB,PRubmann_JPM.1_2018} which was also supported by experimental measurements\cite{ZZhang_Natcom.5_2014}. Recently, a doping concentration-dependent evolution of the exchange mechanism in V and Cr doped  Bi$_2$Te$_3$ was reported. A crossover from van Vleck exchange in the low doping regime to RKKY interaction in the high doping regime was shown to occur experimentally\cite{FWang_NanoLett.23_2023}. However, most of the recent theoretical studies show that the calculated long-ranged exchange interaction strength remains unaffected/insensitive to SOC when it is switched on or off\cite{Jkim_PRB.96_2017, Jkim_PRB.97_2018,TPeixoto_npjQM.5_2020}. Hence, the importance of van Vleck mechanism for a long-ranged interaction becomes questionable. Instead, it was shown for V and Cr doped TIs, that the long-ranged interaction between the doped magnetic atoms is mediated by the polarized $p$-orbital network of the host\cite{Jkim_PRB.96_2017, Jkim_PRB.97_2018}. This polarization occurs due to the $p-d$ hybridization between magnetic dopants and the host atoms\cite{TPeixoto_npjQM.5_2020}. Now if we consider the  short-ranged exchange interactions,  well-known exchange mechanisms like Superexchange and double exchange\cite{Jkim_PRB.92_2015, PRubmann_JPM.1_2018,TPeixoto_npjQM.5_2020} was suggested to play the main role. For example,  AFM superexchange was suggested in Fe-doped Bi$_2$Se$_3$\cite{Jkim_PRB.92_2015} and Mn-doped Bi$_2$Te$_3$\cite{PRubmann_JPM.1_2018}. FM superexchange was reported in Cr-doped Sb$_2$Te$_3$\cite{TPeixoto_npjQM.5_2020}. FM double exchange was suggested to occur in Co-doped Bi$_2$Te$_3$\cite{PRubmann_JPM.1_2018} and V-doped Sb$_2$Te$_3$\cite{TPeixoto_npjQM.5_2020}.  In this work, we take a different approach to investigate the different types of exchange mechanisms that result depending on the dopant type and their short-range/long-range properties. We take advantage of the local octahedral symmetry of the doped transition metal atoms and calculate orbital decomposed exchange interactions using the Magnetic Force Theorem (MFT) to efficiently identify the exchange mechanisms.  For our doped systems, we consider the doping of transition metal (TM) atoms (V, Cr, Mn, and Fe) in one of the Bi layers within a single quintuple layer of Bi$_2$Se$_3$. Such doping creates a hypothetical 2D-like magnet inside a layered non-magnetic topological insulator with possible 2-dimensional magnetic exchange interaction between the TM atoms. However, even this simple 2D magnetic system can incorporate all possible exchange mechanisms considered so far in the literature for explaining the magnetic properties in such systems. We then try to manipulate the electronic structure using local correlation effects in our calculation and measure the changes in the orbital components of the exchange parameters. This helps us to check how sensitive they are to these changes and how their nature could change from one type to another depending on the dopant type and doping concentration.

\section{Methodology}
\subsection{Structural Optimization}
For structural optimization of our systems, the electronic structure was calculated using a projected augmented wave (PAW) method~\cite{blochl_PhysRevB.50.17953_1994,kresse_PhysRevB.59.1758_1999} as implemented in the Vienna ab-initio simulation package 
(VASP)~\cite{kresse_PhysRevB.47.558_1993,kresse_PhysRevB.49.14251_1994,kresse_PhysRevB.54.11169_1996,kresse_cms.6.15_1996}. The local density approximation (LDA)~\cite{dirac_chembridge.26.376_1930,ceperley_PhysRevLett.45.566_1980,perdew_PhysRevB.23.5048_1981} and  generalized gradient approximation (GGA)~\cite{perdew_PhysRevLett.77.3865_1996} in the Perdew-Burke-Ernzerhof (PBE) realization~\cite{perdew_PhysRevLett.77.3865_1996,perdew_PhysRevLett.78.1396_1997} was considered for the exchange-correlation functional. For our layered system the nonlocal, weak van der
Waals (vdW) interactions become important for correct structural properties. These interactions also lead to a contraction of the unit cell volume, correcting the
overestimation in general done by GGA. Hence, we also consider the DFT-D2 method of Grimme~\cite{grimme_jcc.27.15_2006} to include the dispersive interactions in our calculations which are also mainly used in other studies. Relativistic effects also become important due to the presence of heavy elements like Bi. Hence, we also consider the spin-orbit coupling (SOC) effects in our calculations as implemented in the PAW methods in VASP~\cite{steiner_PhysRevB.93.224425_2016}. 
The unit cell volume, shape, and internal positions of all the atoms were optimized for a  minimum energy configuration until the forces were less than 10$^{-3}$ eV/\r{A}. A plane wave energy cutoff of 700 eV was used in all such calculations for primitive and supercells. An optimized gamma-centered
Monkhorst-Pack mesh of 18$\times$18$\times$18 {\bf{k}}-points was used for the Rhombohedral primitive cell. For the conventional three-formula unit(FU) hexagonal cell a k-mesh of  16$\times$16$\times$6 was used. Whenever cell doubling was done, the k-mesh was reduced accordingly.
In our magnetically doped systems, the presence of the transition metal (TM) atoms demands the inclusion of the local correlation effects due to the localized nature of the $3d$ states. To see the effect on the structural properties we also performed one set of GGA+$U$ calculations. This term was then treated in the Hartree-Fock approximation, as done routinely via the DFT+$U$ method~\cite{anisimov_jpcm.9.48_1997,kotliar_RevModPhys.78.865_2006}. In VASP, we employed the rotationally invariant formulation proposed by Liechtenstein {\it{et al.}}~\cite{liechtenstein_PhysRevB.52.R5467_1995}. The Coulomb interaction parameters $U$ and $J$ were chosen based on previous studies~\cite{MFIslam_PRB.97_2018, Kyang_PRB.101_2020} as 4.0 eV and 0.9 eV respectively. 

\subsection{Electronic structure and Magnetic exchange interaction}

Following structural optimization, we conducted electronic and magnetic exchange calculations employing the Full Potential Linear Muffin-Tin Orbital (FPLMTO) method, as implemented in the Relativistic Spin Polarized Toolkit (RSPt)\cite{wills2010full,RSPt}.
For our electronic structure calculations, we used  GGA-PBE, LDA and LDA+$U$ exchange-correlation functionals. Our calculations encompassed two sets of basis functions, covering both valence and semi-core states. These states were specifically constructed from the 6s, 6p, and 5d orbitals for Bi and the 4s, 4p, and 3d orbitals for Se respectively. We chose kinetic tails energy set as -0.1, -2.3, and 1.5 Ry.
In the context of Brillouin zone sampling, we used a gamma-centered Monkhorst-Pack mesh of 14$\times$14$\times$4 {\bf{k}}-points for the conventional three-formula unit(FU) hexagonal cell of both undoped and doped Bi$_2$Se$_3$ systems. Upon introducing the SOC, we increased the \textbf{k}-point sampling to 20$\times$20$\times$6 and   26$\times$26$\times$8 for V, Cr and Mn, Fe doped systems respectively, to achieve precise energy convergence. Furthermore, while doubling the unit cell size the adjustments to the k-mesh were made accordingly.
The LDA+$U$ correction was performed on 100\% doped Bi$_2$Se$_3$ systems using the spin and orbital rotationally invariant formulation provided in the references \cite{PhysRevB_76,GRANAS2012295} using muffin-tin heads as the local basis. The fully localised limit was used as a double counting correction \cite{RevModPhys_78,Vladimir_1997} and other computational settings were kept the same as before.
\\
 The inter-atomic/site magnetic exchange interaction was determined through the application of the magnetic force theorem (MFT)\cite{LIECHTENSTEIN198765,Lichtenstein_PhysRevB_2000}. For this, the $ab-initio$ Kohn-Sham Hamiltonian or the DFT Hamiltonian is mapped onto an effective Heisenberg Hamiltonian with classical spins of the following form\cite{LIECHTENSTEIN198765,Igor_PhysRevB_2015}. 
 
\begin{equation}
 {H} = - \sum_{i\neq j} J_{ij}\vec{e_{i}}\cdot\vec{e_{j}}
 \label{eqn1}
\end{equation}

 Here $(i,j)$ are the indices for the magnetic sites in the system, $\vec{e_i}$, $\vec{e_j}$ are the unit vectors along the spin direction at sites $i$, and $j$ respectively. $J_{ij}$ is the exchange interaction between the two spins at sites $i$ and $j$. In the DFT+DMFT approach in RSPt, $J_{ij}$ has a generalized expression as given below~\cite{Igor_PhysRevB_2015}.

\begin{equation}
J_{ij}=\frac{T}{4} \sum_{n} \operatorname{Tr}\left[\hat{\Delta}_{i}\left(i \omega_{n}\right)\hat{G}_{ij}^{\uparrow}\left(i \omega_{n}\right) \hat{\Delta}_{j}\left(i \omega_{n}\right) \hat{G}_{ji}^{\downarrow}\left(i \omega_{n}\right)\right]
 \label{eqn2}
\end{equation}

Here $T$ is the temperature, and $\hat\Delta$ is the onsite exchange potential giving the exchange splitting at sites $i$, $j$. $\hat{G}_{ij}^{\sigma}$ is the intersite Green's function projected over spin $\sigma$ that can have values $\{\uparrow , \downarrow\}$ and $\omega_n$ is the $n^{th}$ fermionic Matsubara frequency. All the terms in the above expression are matrices in orbital and spin space with the trace running over the orbital degrees of freedom. The summation is over Matsubara frequencies ($\omega_n$). In the presence of strong spin-orbit coupling (SOC), the spin is no longer independent and gets coupled with the orbital degrees of freedom. Due to this the inter-site exchange also becomes anisotropic and a generalized effective Heisenberg Hamiltonian of the following form is considered\cite{KvashninPhysRevB_2020}. 

\begin{eqnarray}
  {H} = - \sum_{i\neq j}   J_{ij}\vec{e_{i}}\cdot\vec{e_{j}} -  \sum_{i\neq j}  \mathbf{e}_{i}^T \mathbb{J}_{ij}^S \mathbf{e}_{j} -  \sum_{i\neq j}  \vec{D}_{ij}\cdot(\vec{e_{i}}\times \vec{e_{j}}) 
  \label{eqn3}
\end{eqnarray}

Where the first term is the isotropic exchange interaction defined in Eqn.~\ref{eqn1}, the second term is the symmetric anisotropic exchange with $\mathbb{J}^S_{ij}$ as the symmetric anisotropic exchange tensor in the form of a $3\times3$ matrix. The third term is the antisymmetric exchange interaction that incorporates Dzyaloshinskii–Moriya(DM) vector $\vec D_{ij}$\cite{Moriya_PhysRev_1960,DZYALOSHINSKY1958241}. This interaction is expressed in the form of a scalar triple product. 

To ensure precise convergence in our analysis of exchange parameters, we increased the k-mesh to 70$\times$70$\times$24 for our magnetic exchange interaction calculations. While doubling the cell the corresponding k-mesh was reduced accordingly.
    

\section{Results and Discussions}

\subsection{Structural optimization with VASP}

\subsubsection{The best method for pristine Bi$_2$Se$_3$}

We started by searching for a method for structural optimization that will give the best result for pristine Bi$_2$Se$_3$ close to experimental results. We shall then use that method for the structural optimization of our doped magnetic systems. We considered both GGA-PBE as well as LDA for the exchange-correlation functional, with vdW and SOC corrections (see methodology section for details) which are important for these layered materials containing heavy ions. We considered all possible combinations to fully optimize the structure and extract the values associated with some of the structural parameters given in Table~\ref{Table1} along with the experimental values. We observe that GGA-PBE alone overestimates the lattice parameters and bond lengths. The inclusion of the vdW-corrections corrects this overestimation and the parameter values become close to the experimental values. The inclusion of SOC is only able to induce smaller modifications. LDA on the other hand underestimates the structural degrees of freedom and the inclusion of SOC further enhances this effect. The DFT-D2 method for vdW corrections that we use is not compatible with LDA functional as the functional parametrization is incomplete. In an exploratory effort, we forcefully used vdW correction with LDA using the GGA-PBE parameter value (the last two rows in Table~\ref{Table1}) which gives erroneous results with a large underestimation of the structural parameters and shall not be used for these systems. The values in bold and italics in each column are the first and second closest ones to the experimental values~\cite{perez_ing_chem.38.9_1999} that we obtain from our calculations. We found that PBE with vdW (DFT-D2) and SOC corrections gives the best result. The next best method shall be PBE with only vdW corrections. We shall note that the SOC-induced corrections to the structural parameters are not so huge and hence PBE+vdW (DFT-D2) method can also be used for the structural optimization of larger systems to save computation time.

\FloatBarrier
\begin{table}[h]
\caption{Table showing different parameter values (lattice constants \textit{\textbf{a}} and \textit{\textbf{c}}, cell volume \textbf{V}, Se-Bi bond lengths, and the inter-layer vdW gap) in a 
3FU hexagonal bulk Bi$_2$Se$_3$ obtained from various treatments of exchange-correlation functionals with SOC and vdW-corrections in DFT and
the corresponding experimental values ($^a$Ref.~\cite{nakajima_jpcs.24.3_1963}, $^b$Ref.~\cite{horak_jpcs.51.12_1990}, $^c$Ref.~\cite{perez_ing_chem.38.9_1999}). In each column, the values in bold and italics are the first and second closest ones to the experimental values (Expt.$^c$).}
\label{Table1}
\begin{tabular}{|l|l|l|l|l|l|l|}
\hline
\textbf{} & \textit{\textbf{a}} (\r{A}) & \textit{\textbf{c}} (\r{A}) & \textbf{V} (\r{A}$^3$) & \textbf{Se$_2$-Bi} (\r{A}) & \textbf{Se$_1$-Bi} (\r{A}) & \textbf{vdW gap} (\r{A}) \\ \hline
Expt.$^a$ & 4.143 & 28.636 & 425.670 & 3.03 & 2.89 & 2.568 \\ \hline
Expt.$^b$ & 4.138 & 28.624 & 424.568 & 3.04 & 2.87 & 2.590 \\ \hline
Expt.$^c$ & 4.135 & 28.615 & 423.819 & 3.06 & 2.86 & 2.530 \\ \hline
PBE & 4.190 & 31.410 & 477.590 & 3.10 & 2.88 & 3.441 \\ \hline
PBE+SOC & 4.201 & 30.663 & 468.574 & 3.11 & 2.89 & 3.168 \\ \hline
PBE+vdW(D2) & \textit{4.125} & 28.975 & \textit{427.000} & \textit{3.06} & \textbf{2.86} & 2.664 \\ \hline
PBE+vdW(D2)+SOC & \textbf{4.135} & \textbf{28.713} & \textbf{425.064} & \textbf{3.06} & \textit{2.87} & \textbf{2.555} \\ \hline
PBE+vdW(D3) & 4.175 & \textit{28.886} & 436.127 & 3.09 & 2.87 & \textit{2.649} \\ \hline
LDA & 4.107 & 27.949 & 408.372 & 3.04 & 2.84 & 2.385 \\ \hline
LDA+SOC & 4.116 & 27.714 & 406.654 & 3.04 & 2.85 & 2.277 \\ \hline
LDA+vdW(D2) & 4.057 & 27.092 & 386.238 & 2.99 & 2.82 & 2.173 \\ \hline
LDA+vdW(D2)+SOC & 4.062 & 26.920 & 384.667 & 2.99 & 2.83 & 2.084 \\ \hline
\end{tabular}
\end{table}
\FloatBarrier

\subsubsection{Modelling our doped Bi$_2$Se$_3$ systems}

\begin{figure}[h]
\includegraphics[scale=0.5]{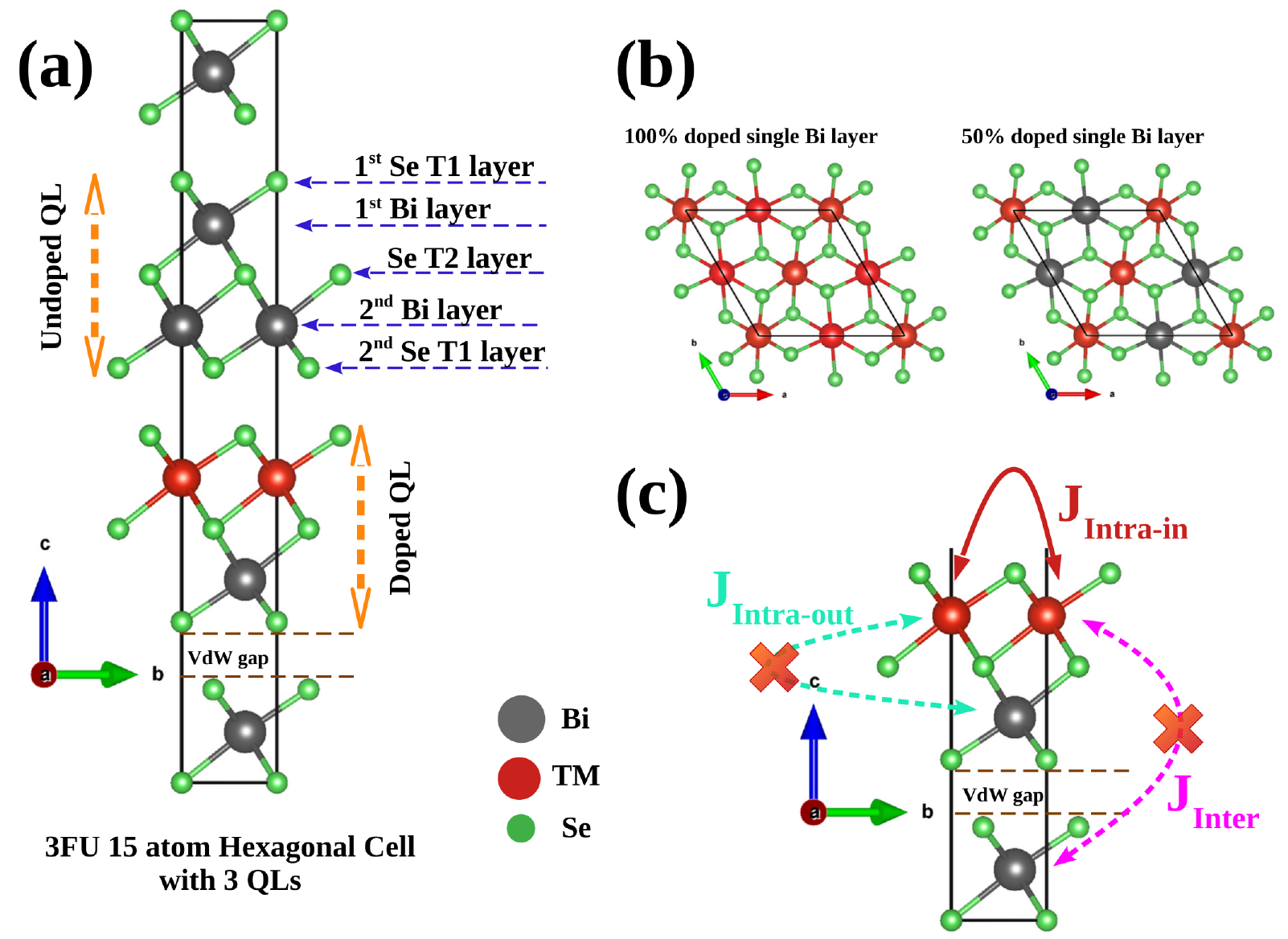}
\caption{(a) 3FU hexagonal Bi$_2$Se$_3$ unit cell with a doped Quintuple layer (QL). Different atomic layers/planes in the undoped QL are also shown. Bi, Se and transition metal (TM) atoms are indicated by grey, green, and red spheres respectively. (b) A bird's eye view of the single-doped Bi-plane for 100\% and 50\% doping cases. (c) The possible inter-atomic exchange mechanisms in this type of stacked QL system with realistic doping at the cationic site. Details in the main text.}
\label{Struc_dop-tech-crop}
\end{figure}
\FloatBarrier

In our doped systems, doping of transition metal (TM) atom is introduced within a single Quintuple Layer (QL) and only in the first Bi atomic layer of that QL as shown schematically in Fig.~\ref{Struc_dop-tech-crop}(a). The other QLs in the system remain undoped. We have considered the TM atoms V, Cr, Mn, and Fe as the dopant atoms. As the planar doping concentration decreases, the average separation between the TM atoms increases as shown in Fig.~\ref{Struc_dop-tech-crop}(b). Such doping creates
a hypothetical 2D-like magnet inside a layered non-magnetic topological insulator with the only possible intra-layer in-plane ($J_{Intra-in}$) magnetic exchange interaction between the TM atoms in the same Bi layer of a QL as shown schematically in Fig.~\ref{Struc_dop-tech-crop}(c). For realistic doping, there shall be TM atoms in all the stacked QLs and shall act as a layered magnetic vdW material with inter-layer ($J_{Inter}$) interactions between TM atoms in two adjacent QL, as well as intra-layer out-plane ($J_{Intra-out}$) interactions between TM atoms in two separate Bi layers of a single QL. In our system, only the  $J_{Intra-in}$ is present and we call it $J_{ij}$ from now. The reason for considering a model 2D magnetic system like this is that we want to study the origin and nature of the in-plane magnetic exchange interaction between the TM atoms, and their dependence as a function of in-plane doping concentrations in an ideal environment in the absence of the inter-layer and out-of-plane intra-layer interactions. The in-plane intra-layer interaction that we consider here is supposed to be the easiest one to model, analyze, and understand, but can incorporate all the possible exchange mechanisms that usually occur in these systems with more realistic doping. Hence a detailed understanding of the magnetic interactions even in this ideal system can help us understand a realistic doped system in a better manner. This is going to be important for both theoretical and technological purposes.  Another reason for this restriction is that
here we intend to calculate the magnetic exchange using the Magnetic Force Theorem (MFT) and model our magnetically doped system using the supercell approach which shall
take into account local distortions, and structural anisotropy (as a result of doping) in a
real manner not possible with a mean-field approach like Coherent Potential approximation(CPA). This will allow us to check for the long-range nature of the magnetic exchange with doping concentrations in the presence of local structural distortions. Modelling a real doped system via a supercell approach and then applying MFT shall require a large system size impossible to treat within DFT calculations.  Moreover, in this system,
the quintuple layers shall be weakly coupled via vdW interactions and it should be possible to isolate
them as single quasi-2D layers. Our study shall become more reasonable and comparable for such isolated
2D layers. 

\subsubsection{Structural optimization method for doped Bi$_2$Se$_3$}

For the structural optimization of our doped systems, we consider the results that we got for pristine Bi$_2$Se$_3$ and choose GGA-PBE+vdW (DFT-D2) for the exchange-correlation functional, SOC for the heavy atom Bi, and local correlation effect (U) for the transition metal d-states. We then apply these methodological schemes in all possible ways to our TM-doped system with 100\% doping of a single Bi-layer. The outcome of this structural optimization is very similar for all the TM-doped systems. Here we report the results for the Cr case in Table.\ref{Table2} as a reference. We see that the effect of U is to increase the local Cr-Se bonds by about 0.03 \AA, without affecting the Bi-Se bonds both in the doped and undoped QLs. This also increases the lattice parameters and system volume slightly. The inclusion of SOC on the other hand does not affect the Cr-Se bonds. It slightly increases the Bi-Se bonds and contracts the cell volume. The additional structural changes induced by U, SOC, and U+SOC together have a negligible effect on the electronic structure. Electronic structure calculation performed with the same computational settings and taking the four structures reported in Table.~\ref{Table2} are almost identical (See section I of the SM for details). The major structural changes as a result of doping are well captured within the GGA-PBE+vdW approach. Hence,  to reduce the computational cost, we use the same for structural optimization for all our doped systems and include SOC or U only during  the electronic structure calculation with these structures in RSPt as required.

\begin{table}[]
\caption{Calculated lattice parameters, cell volume, cation-anion bond lengths, and magnetic moments using different methodological schemes for the Cr-doped system.}
\label{Table2}
\begin{tabular}{|c|c|c|c|ccccc|ccc|cc|}
\hline
\multirow{2}{*}{\textbf{Method}} & \multirow{2}{*}{\textbf{\begin{tabular}[c]{@{}c@{}}a\\ (\AA)\end{tabular}}} & \multirow{2}{*}{\textbf{\begin{tabular}[c]{@{}c@{}}c\\ (\AA)\end{tabular}}} & \multirow{2}{*}{\textbf{\begin{tabular}[c]{@{}c@{}}Vol\\ (\AA)\end{tabular}}} & \multicolumn{5}{c|}{\textbf{\begin{tabular}[c]{@{}c@{}}n.n bonds in \\ doped QL\end{tabular}}} & \multicolumn{3}{c|}{\textbf{\begin{tabular}[c]{@{}c@{}}n.n bonds in \\ undoped QL\end{tabular}}} & \multicolumn{2}{c|}{\textbf{\begin{tabular}[c]{@{}c@{}}Moment \\ ($\mu_B$)\end{tabular}}} \\ \cline{5-14} 
 &  &  &  & \multicolumn{1}{c|}{\textbf{\begin{tabular}[c]{@{}c@{}}M-\\ Se2\end{tabular}}} & \multicolumn{1}{c|}{\textbf{\begin{tabular}[c]{@{}c@{}}M-\\ Se1\end{tabular}}} & \multicolumn{1}{c|}{\textbf{\begin{tabular}[c]{@{}c@{}}Bi-\\ Se2\end{tabular}}} & \multicolumn{1}{c|}{\textbf{\begin{tabular}[c]{@{}c@{}}Bi-\\ Se1\end{tabular}}} & \textbf{\begin{tabular}[c]{@{}c@{}}vdW \\ gap\end{tabular}} & \multicolumn{1}{c|}{\textbf{\begin{tabular}[c]{@{}c@{}}Bi-\\ Se2\end{tabular}}} & \multicolumn{1}{c|}{\textbf{\begin{tabular}[c]{@{}c@{}}Bi-\\ Se1\end{tabular}}} & \textbf{\begin{tabular}[c]{@{}c@{}}vdW\\ gap\end{tabular}} & \multicolumn{1}{c|}{\textbf{M}} & \textbf{Cell} \\ \hline
PBE+vdW & 4.05 & 28.53 & 405.32 & \multicolumn{1}{c|}{2.68} & \multicolumn{1}{c|}{2.61} & \multicolumn{1}{c|}{3.09} & \multicolumn{1}{c|}{2.83} & 2.73 & \multicolumn{1}{c|}{3.05} & \multicolumn{1}{c|}{2.84} & 2.72 & \multicolumn{1}{c|}{3.23} & 3.00 \\ \hline
PBE+vdW+U & 4.06 & 28.57 & 407.79 & \multicolumn{1}{c|}{2.71} & \multicolumn{1}{c|}{2.64} & \multicolumn{1}{c|}{3.09} & \multicolumn{1}{c|}{2.83} & 2.71 & \multicolumn{1}{c|}{3.06} & \multicolumn{1}{c|}{2.84} & 2.71 & \multicolumn{1}{c|}{3.52} & 3.00 \\ \hline
\begin{tabular}[c]{@{}c@{}}PBE+vdW+SOC\end{tabular} & 4.06 & 28.35 & 403.87 & \multicolumn{1}{c|}{2.68} & \multicolumn{1}{c|}{2.61} & \multicolumn{1}{c|}{3.11} & \multicolumn{1}{c|}{2.83} & 2.67 & \multicolumn{1}{c|}{3.06} & \multicolumn{1}{c|}{2.85} & 2.62 & \multicolumn{1}{c|}{\begin{tabular}[c]{@{}c@{}}3.23 $\hat{\textbf{z}}$\end{tabular}} & \begin{tabular}[c]{@{}c@{}}3.00 $\hat{\textbf{z}}$\end{tabular} \\ \hline
\begin{tabular}[c]{@{}c@{}}PBE+vdW+SOC+ U\end{tabular} & 4.07 & 28.37 & 406.28 & \multicolumn{1}{c|}{2.71} & \multicolumn{1}{c|}{2.64} & \multicolumn{1}{c|}{3.11} & \multicolumn{1}{c|}{2.84} & 2.66 & \multicolumn{1}{c|}{3.06} & \multicolumn{1}{c|}{2.85} & 2.61 & \multicolumn{1}{c|}{\begin{tabular}[c]{@{}c@{}}3.52 $\hat{\textbf{z}}$\end{tabular}} & \begin{tabular}[c]{@{}c@{}}3.23 $\hat{\textbf{z}}$\end{tabular} \\ \hline
\end{tabular}
\end{table}

\subsection{Electronic and Magnetic properties with RSPt}

For the electronic structure calculation in RSPt, we tested three approximations for the exchange-correlation functionals, LDA, GGA-PBE, and GGA-AM05 to calculate the electronic structure of pristine bulk Bi$_2$Se$_3$. We found that LDA gives the best description of the calculated band structure in agreement with the literature and VASP calculated band structure using GGA-PBE. Hence, we choose LDA functional for all the calculations of electronic structure and exchange interactions of our TM doped Bi$_2$Se$_3$ systems. The details can be found in Section II of the SM. 

\subsubsection{The 100\% doped case}

We start with a brief description of the structural changes as a result of the doping that we got from the VASP calculations. The details are tabulated in Table~\ref{Table3} for all the systems. The most visible effect of doping is the decrease in the lattice parameters and unit cell volume as a result of the smaller size of the TM atoms (M) compared to Bi. The M-Se bonds are also smaller compared to the Bi-Se bonds. Bi-Se bonds in the undoped QLs remain unaffected due to this type of doping maintaining their pristine local chemical properties. Going from V to Fe, the M-Se bonds do not show any systematic change. This is because it depends on both the atomic size and bonding interaction with the Se atoms. The only parameter that systematically decreases from V to Fe is the vdW gap between the doped and adjacent undoped QLs. It decreases from 2.83 \r{A} in V to 2.59 \r{A} in Fe. As a result, the electronic and magnetic properties of the Mn and especially Fe-doped layers are expected to be more  affected by the neighbouring undoped QLs compared to V and Cr-doped layers.

\begin{table}[]
\caption{Variation in the lattice parameters, cell volume, and cation-anion bond lengths with the doped TM-atoms from PBE+vdW VASP calculations.}
\label{Table3}
\begin{tabular}{|c|c|c|c|ccccc|ccc|}
\hline
\multirow{2}{*}{\textbf{\begin{tabular}[c]{@{}c@{}}TM \\ atom \\ (M)\end{tabular}}} &
  \multirow{2}{*}{\textbf{\begin{tabular}[c]{@{}c@{}}a\\ (\AA)\end{tabular}}} &
  \multirow{2}{*}{\textbf{\begin{tabular}[c]{@{}c@{}}c\\ (\AA)\end{tabular}}} &
  \multirow{2}{*}{\textbf{\begin{tabular}[c]{@{}c@{}}Vol\\ (\AA$^3$)\end{tabular}}} &
  \multicolumn{5}{c|}{\textbf{\begin{tabular}[c]{@{}c@{}}n.n bonds in\\ doped QL (\AA)\end{tabular}}} &
  \multicolumn{3}{c|}{\textbf{\begin{tabular}[c]{@{}c@{}}n.n bonds in \\ undoped QL (\AA)\end{tabular}}} \\ \cline{5-12} 
 &
   &
   &
   &
  \multicolumn{1}{c|}{\textbf{\begin{tabular}[c]{@{}c@{}}M-\\ Se2\end{tabular}}} &
  \multicolumn{1}{c|}{\textbf{\begin{tabular}[c]{@{}c@{}}M-\\ Se1\end{tabular}}} &
  \multicolumn{1}{c|}{\textbf{\begin{tabular}[c]{@{}c@{}}Bi-\\ Se2\end{tabular}}} &
  \multicolumn{1}{c|}{\textbf{\begin{tabular}[c]{@{}c@{}}Bi-\\ Se1\end{tabular}}} &
  \textbf{\begin{tabular}[c]{@{}c@{}}vdW\\ gap\end{tabular}} &
  \multicolumn{1}{c|}{\textbf{\begin{tabular}[c]{@{}c@{}}Bi-\\ Se2\end{tabular}}} &
  \multicolumn{1}{c|}{\textbf{\begin{tabular}[c]{@{}c@{}}Bi-\\ Se1\end{tabular}}} &
  \textbf{\begin{tabular}[c]{@{}c@{}}vdW\\ gap\end{tabular}} \\ \hline
V &
  4.06 &
  28.59 &
  407.42 &
  \multicolumn{1}{c|}{2.77} &
  \multicolumn{1}{c|}{2.54} &
  \multicolumn{1}{c|}{3.08} &
  \multicolumn{1}{c|}{2.83} &
  2.83 &
  \multicolumn{1}{c|}{3.05} &
  \multicolumn{1}{c|}{2.84} &
  2.72 \\ \hline
Cr &
  4.05 &
  28.53 &
  405.32 &
  \multicolumn{1}{c|}{2.68} &
  \multicolumn{1}{c|}{2.61} &
  \multicolumn{1}{c|}{3.09} &
  \multicolumn{1}{c|}{2.83} &
  2.73 &
  \multicolumn{1}{c|}{3.05} &
  \multicolumn{1}{c|}{2.84} &
  2.72 \\ \hline
Mn &
  4.06 &
  28.55 &
  406.87 &
  \multicolumn{1}{c|}{2.75} &
  \multicolumn{1}{c|}{2.59} &
  \multicolumn{1}{c|}{3.09} &
  \multicolumn{1}{c|}{2.84} &
  2.67 &
  \multicolumn{1}{c|}{3.05} &
  \multicolumn{1}{c|}{2.84} &
  2.72 \\ \hline
Fe &
  4.05 &
  28.29 &
  401.06 &
  \multicolumn{1}{c|}{2.77} &
  \multicolumn{1}{c|}{2.55} &
  \multicolumn{1}{c|}{3.04} &
  \multicolumn{1}{c|}{2.84} &
  2.59 &
  \multicolumn{1}{c|}{3.04} &
  \multicolumn{1}{c|}{2.84} &
  2.67 \\ \hline
\end{tabular}
\end{table}

\FloatBarrier
\begin{table}[]
\caption{Orbital resolved nearest neighbour exchange parameter, spin-resolved  $3d$ charge, and transition metal (TM) magnetic moment for all the TM-doped systems with and without local correlation effects.}
\label{Table4}
\begin{tabular}{|c|c|cccc|ccc|c|}
\hline
\multirow{2}{*}{\begin{tabular}[c]{@{}c@{}}TM \\ atom\end{tabular}} &
  \multirow{2}{*}{U, J (eV)} &
  \multicolumn{4}{c|}{\textbf{\begin{tabular}[c]{@{}c@{}}Orbital-resolved n.n \\ exchange parameters\\ (mRy)\end{tabular}}} &
  \multicolumn{3}{c|}{\textbf{\begin{tabular}[c]{@{}c@{}}Spin resolved \\ d-orbital charge\\ (electron unit)\end{tabular}}} &
  \multirow{2}{*}{\textbf{\begin{tabular}[c]{@{}c@{}}$\mu_{TM}$\\ ($\mu_{B}$)\end{tabular}}} \\ \cline{3-9}
 &
   &
  \multicolumn{1}{c|}{e$_g$ - e$_g$} &
  \multicolumn{1}{c|}{\begin{tabular}[c]{@{}c@{}}T$_{2g}$ - \\ T$_{2g}$\end{tabular}} &
  \multicolumn{1}{c|}{\begin{tabular}[c]{@{}c@{}}e$_g$ - \\ T$_{2g}$\end{tabular}} &
  Total &
  \multicolumn{1}{c|}{Q$_d$-up} &
  \multicolumn{1}{c|}{Q$_d$-dn} &
  $\Delta$Q &
   \\ \hline
\multirow{3}{*}{V} &
  0.0, 0.0 &
  \multicolumn{1}{c|}{-0.024} &
  \multicolumn{1}{c|}{0.670} &
  \multicolumn{1}{c|}{0.371} &
  1.017 &
  \multicolumn{1}{c|}{2.38} &
  \multicolumn{1}{c|}{0.45} &
  1.93 &
  1.96 \\ \cline{2-10} 
 &
  3.0, 0.9 &
  \multicolumn{1}{c|}{-0.073} &
  \multicolumn{1}{c|}{-0.064} &
  \multicolumn{1}{c|}{0.291} &
  0.153 &
  \multicolumn{1}{c|}{2.32} &
  \multicolumn{1}{c|}{0.44} &
  1.88 &
  1.91 \\ \cline{2-10} 
 &
  4.0, 0.9 &
  \multicolumn{1}{c|}{-0.059} &
  \multicolumn{1}{c|}{-0.056} &
  \multicolumn{1}{c|}{0.246} &
  0.130 &
  \multicolumn{1}{c|}{2.32} &
  \multicolumn{1}{c|}{0.42} &
  1.90 &
  1.93 \\ \hline
\multirow{2}{*}{Cr} &
  0.0, 0.0 &
  \multicolumn{1}{c|}{0.000} &
  \multicolumn{1}{c|}{-0.076} &
  \multicolumn{1}{c|}{0.836} &
  0.760 &
  \multicolumn{1}{c|}{3.42} &
  \multicolumn{1}{c|}{0.45} &
  2.97 &
  3.03 \\ \cline{2-10} 
 &
  3.0, 0.9 &
  \multicolumn{1}{c|}{-0.013} &
  \multicolumn{1}{c|}{-0.017} &
  \multicolumn{1}{c|}{0.557} &
  0.528 &
  \multicolumn{1}{c|}{3.45} &
  \multicolumn{1}{c|}{0.40} &
  3.05 &
  3.10 \\ \hline
\multirow{2}{*}{Mn} &
  0.0, 0.0 &
  \multicolumn{1}{c|}{0.391} &
  \multicolumn{1}{c|}{-0.100} &
  \multicolumn{1}{c|}{-0.271} &
  0.019 &
  \multicolumn{1}{c|}{4.28} &
  \multicolumn{1}{c|}{0.58} &
  3.70 &
  3.77 \\ \cline{2-10} 
 &
  3.0, 0.9 &
  \multicolumn{1}{c|}{0.359} &
  \multicolumn{1}{c|}{-0.011} &
  \multicolumn{1}{c|}{0.004} &
  0.353 &
  \multicolumn{1}{c|}{4.42} &
  \multicolumn{1}{c|}{0.42} &
  4.00 &
  4.07 \\ \hline
\multirow{2}{*}{Fe} &
  0.0, 0.0 &
  \multicolumn{1}{c|}{1.505} &
  \multicolumn{1}{c|}{0.777} &
  \multicolumn{1}{c|}{-5.115} &
  -2.832 &
  \multicolumn{1}{c|}{4.41} &
  \multicolumn{1}{c|}{1.53} &
  2.88 &
  2.92 \\ \cline{2-10} 
 &
  3.0, 0.9 &
  \multicolumn{1}{c|}{0.004} &
  \multicolumn{1}{c|}{-1.220} &
  \multicolumn{1}{c|}{-0.448} &
  -1.665 &
  \multicolumn{1}{c|}{4.61} &
  \multicolumn{1}{c|}{1.27} &
  3.34 &
  3.39 \\ \hline
\end{tabular}
\end{table}
\FloatBarrier

\begin{figure}[h]
\includegraphics[scale=0.55]{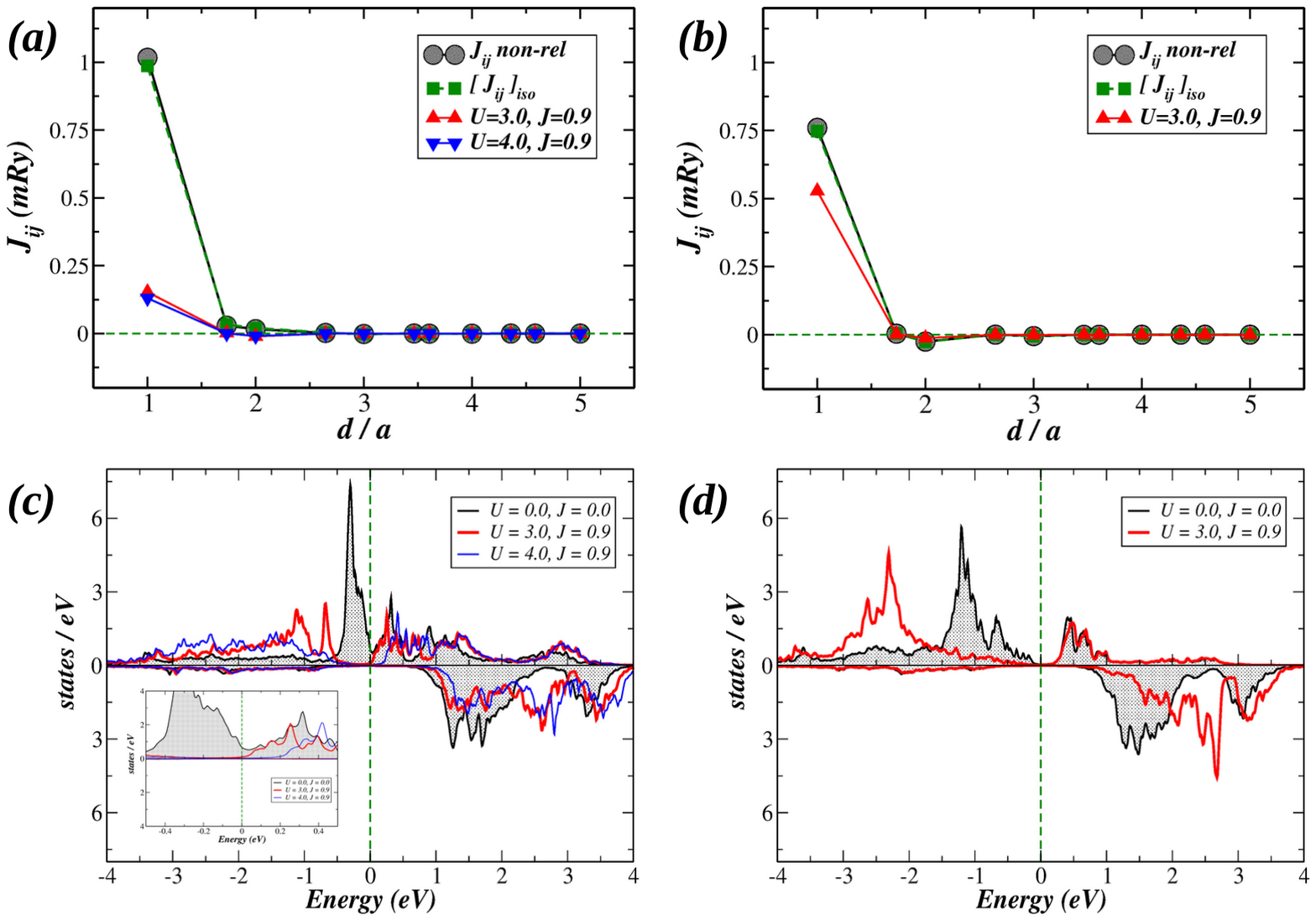}
\caption{TOP panel: Comparison between the calculated non-relativistic (with and without local correlation) magnetic exchange parameters and isotropic part of the magnetic exchange tensor (without local correlation) in the presence of SOC for the (a) V and (b) Cr doped (100\%) cases. BOTTOM panel: (c) pdos corresponding to the V $3d$ states with and without the local correlation effects. The inset shows the states around Fermi energy and the finite contribution for the plain LDA case.  (d) pdos corresponding to the Cr $3d$ states with and without the local correlation effect. pdos = projected density of states.}
\label{Bulk-100per_V-Cr}
\end{figure}
\FloatBarrier

For the electronic structure and magnetic properties, we do calculations with and without SOC/local correlation ($U$). The magnetic exchange parameters without SOC are then compared with the isotropic part of the exchange tensor calculated in the presence of SOC to see its role in the magnetic exchange interactions. This comparison is made only in the absence of local correlation ($U$) effects. We start our discussion with the V and Cr systems. The calculated exchange parameters for V and Cr are shown in Fig.~\ref{Bulk-100per_V-Cr} (a) and (b) respectively. We can see that the calculated exchange parameters with (green squares) and without SOC (black spheres) are identical. This suggests that SOC, in this case, is not important in determining the magnetic exchange interactions. This is reasonable and expected in the 100\% doping case where the heavy Bi atoms in the plane are all replaced by TM atoms. This is in agreement with previous studies where heavy Cr doping in (Bi, Sb)$_2$Se$_3$ turns the system into a trivial FM insulator with reduced SOC effects\cite{zhang_science.339_2013,zhao_nature.588_2020}. Next, we focus on the distance dependence of the calculated exchange parameters which are found to be short-ranged with an FM nearest neighbour (n.n) interaction. The second and third neighbour interactions having non-zero values are very small compared to the n.n interaction strength and can be ignored in this case as they are not going to play any role in the magnetic ordering.  To gain a precise understanding of the nature and origin of the n.n interaction, we now take advantage of the octahedral symmetry and have extracted the orbital resolved $e_g-e_g$, $T_{2g}-T_{2g}$, and $e_g-T_{2g}$ components of the n.n exchange which are presented in Table~\ref{Table4}. Additionally, the table includes theoretically obtained local magnetic moment ($\mu_{TM}$) of the TM atoms, $d$-orbital projected charge occupations ($Q_d$) in different spin channels and their differences ($\Delta Q$). The table shows that in the V-doped system without $U$, the major FM contribution to the n.n exchange comes from the  $T_{2g}-T_{2g}$ component (0.67 mRy) followed by the $e_g-T_{2g}$ component (0.37 mRy). A local magnetic moment of 1.96 $\mu_{B}$ suggests a d$^2$ electron configuration in the 3+ oxidation state, which is consistent with the previous literature \cite{verginory_PhysRevB, MFIslam_PRB.97_2018}. Though the $d$-orbital charge occupation in the majority and minority spin channels is more than 2 and 0 due to hybridization with the Se $p$-states, their difference is close to 2 supporting the observed magnetic moment. The local projected density of states (pdos) corresponding to the V $ 3d$ states are shown in Fig.~\ref{Bulk-100per_V-Cr} (c) as a grey-shaded region. Due to the d$^2$ electron configuration, partially filled $T_{2g}$-up states appear as an impurity band in the majority spin channel with a finite contribution at the Fermi level ($E_F$). The empty $e_g$-up states in the majority spin channel are further higher in energy in the conduction band. A virtual hopping from the partially filled $T_{2g}$-up states  to the empty $ e_g$ up states of the nearest neighbour V atoms is possible through Se atoms. A nearly 90$^{\circ}$ angle in this V-Se-V exchange channel can give rise to an FM superexchange following the Goodenough-Kanamori-Anderson(GKA) rules\cite{Wang_2022-GKA}, which was also reported in a very similar system VI$_3$ where V is also with a d$^2$ electron configuration\cite{Kyang_PRB.101_2020}. The virtual hopping process mentioned above involves two component mechanisms. First, a spin parallel electron is transferred from the filled $p$-orbital of the Se atom to the empty $e_g$-up orbital of the first V atom. This results in an energy gain following intra-atomic Hund’s rule leading to an FM V-Se coupling. The second electron from the Se $p$-orbital couples antiferromagnetically with the filled $T_{2g}$ orbital of the second V atom due to a direct exchange between non-orthogonal orbitals. The net $e_g - T_{2g}$ exchange between the two V atoms becomes FM. This suggests that our FM $e_g-T_{2g}$ component comes from a superexchange mechanism.  
Now, a virtual hopping between the partially filled $T_{2g}$-up orbitals of two nearest neighbour V atoms is also possible. This can also result in a $T_{2g}-T_{2g}$ FM superexchange mechanism similar to the case of VI$_3$ system\cite{Kyang_PRB.101_2020}. Although it cannot overweigh the $e_g-T_{2g}$ FM superexchange mechanism as the latter involves sigma bonding between the Se $p$-orbital and the V $e_g$-orbital along the bonding axis \cite{Wang_2022-GKA}. However, the presence of finite $T_{2g}$-up states at $E_F$ (shown in the inset of Fig.~\ref{Bulk-100per_V-Cr}(c) for clarity) gives rise to the possibility of a $T_{2g}-T_{2g}$ FM double exchange mechanism\cite{TJungwirth_RevModPhys.78.809_2006,KSato_RevModPhys.82.1633_2010} that can dominate over superexchange interactions. In order to understand the role of finite impurity-like $T_{2g}$-up states at $E_F$ in the exchange mechanism, we perform an LDA+$U$ calculation with U=3.0 eV and J=0.9 eV with an expectation to remove the $T_{2g}$-up states from  $E_F$ and then measure the changes in the exchange components. Here, we use the local correlation as a tool to perturb the electronic structure and measure the corresponding changes in the exchange parameters. Finding the most accurate electronic structure with the correct value of the correlation parameter is not our aim here. Hence, any other value of U could have been chosen. The corresponding exchange parameters (red up-triangles) and orbital projected $3d$-states (red lines) are shown in Fig.~\ref{Bulk-100per_V-Cr} (a) and (c), respectively.  As expected, introducing a local correlation/Hubbard U drives the system insulating with the filled part of the $T_{2g}$-up states in the majority spin channel pushed further below the Fermi level. As a result, there is a noticeable change in the n.n T$_{2g}$- T$_{2g}$ exchange component, transitioning from the dominant FM (0.67) to a weak AFM (-0.06), which is expected to be superexchange type in this insulating state (Table~\ref{Table4}). This confirms that the dominant n.n  T$_{2g}$- T$_{2g}$ component without local correlation effect comes from a double exchange mechanism due to impurity-like states at $E_F$. With $U$, the $e_g-T_{2g}$ FM superexchange now becomes the dominant mechanism with a value of 0.29 mRy. This value decreases from  0.37 mRy (no U case) due to the increased separation between filled $T_{2g}$-up and empty $e_g$-up states as a result of the local correlation effect but does not die like the $T_{2g} - T_{2g}$ component, again confirming it's superexchange nature. To further explore whether the FM superexchange mechanism remains prevalent, we increased $U$ to 4 eV. This leads to a further increase in the separation between the T$_{2g}$ and e$_g$ states as shown by the blue lines in Fig.~\ref{Bulk-100per_V-Cr} (c). This change in the electronic structure however further slightly decreases the strength of the FM superexchange T$_{2g}$-e$_g$ component from 0.29 to 0.25 mRy (Table~\ref{Table4}). \\

As we go from the V to the Cr, one more electron is added to the $d$-orbital. Consequently, we can anticipate a $d^3$ electron configuration if Cr is in a 3+ oxidation state similar to V. As reported in Table~\ref{Table4}, an occupation of 3.42 and 0.45 $e$ in the majority and minority $3d$-states respectively, along with a local moment of 3.03 $\mu_B$, confirms the $d^3$ configuration of Cr. This again aligns with some previous findings from similar Cr-doped systems\cite{verginory_PhysRevB,MFIslam_PRB.97_2018}.
The local pdos of the Cr-$3d$ orbitals as shown in Fig.~\ref{Bulk-100per_V-Cr}(d), also illustrates a d$^3$ configuration with a filled $T_{2g}$-up in the valence band and an empty $e_g$-up in the conduction band. A large separation between them creates an insulating gap. Unlike the V-doped system, this insulating gap and a filled $T_{2g}$-up eliminates the possibility of a  $T_{2g}-T_{2g}$ FM double exchange mechanism. Component analysis of the n.n exchange as given in Table~\ref{Table4} also supports this showing a weak antiferromagnetic $T_{2g}-T_{2g}$ component. This is again following the GKA rules\cite{Wang_2022-GKA} as described before. As a result, in this case, the virtual hopping between filled $T_{2g}$-up and vacant $e_g$-up gives rise to the dominant $e_g-T_{2g}$ FM superexchange of 0.84 mRy. These observations are again in line with the analogous system CrI$_3$ where Cr is also Cr3+ with a nearly 90$^o$ Cr-I-Cr exchange path\cite{HWang_EPL.114_2016,XFeng_PRB.100_2019}.
Following what we did for the V case, we again do an LDA+$U$ calculation with a U and J value of 3.0 and 0.9 eV on the Cr-$3d$ states, to study the impact on the exchange interactions and electronic structure. The calculated exchange parameters (red triangles) and $3d$ states (red bold lines) are plotted in Fig.~\ref{Bulk-100per_V-Cr}(b) and (d) respectively. A local correlation increases the separation between $T_{2g}$-up and $e_g$-up states as expected leading to an increased band gap. This enhanced $T_{2g}-e_g$ separation results in a decrease in the n.n $e_g-T_{2g}$ superexchange interaction from 0.84 mRy to 0.56 mRy but remains as the dominant mechanism. The n.n $T_{2g}-T_{2g}$ superexchange interaction strength slightly decreases (see Table~\ref{Table4}) and distant neighbour interactions  however remain unaffected(Fig.~\ref{Bulk-100per_V-Cr}(b).

These results show that for V, both double exchange and superexchange could result in a short-ranged FM interaction due to a partially filled $T_{2g}$ band where the double exchange mechanism strongly depends on the states around Fermi energy. For Cr, however, only the superexchange mechanism is possible for the n.n interaction. Our results are in agreement with a recent study\cite{TRFPeixoto_NJPQM.5_2020} where the dependence of the calculated exchange parameter on the position of the Fermi energy is found to be different in V and Cr doped Sb$_2$Te$_3$. They show that for V, a sharp peak followed by a broad ridge appears in $J_{ij}(E)$ when $E_F$ moves from inside the V $T_{2g}$ manifold to in between $T_{2g}$ and $e_g$. The sharp peak has been associated with a double exchange mechanism that does not appear for the Cr system. The broad ridge has been associated with a superexchange mechanism which is present in both V and Cr systems. \\

\begin{figure}[h]
\includegraphics[scale=0.55]{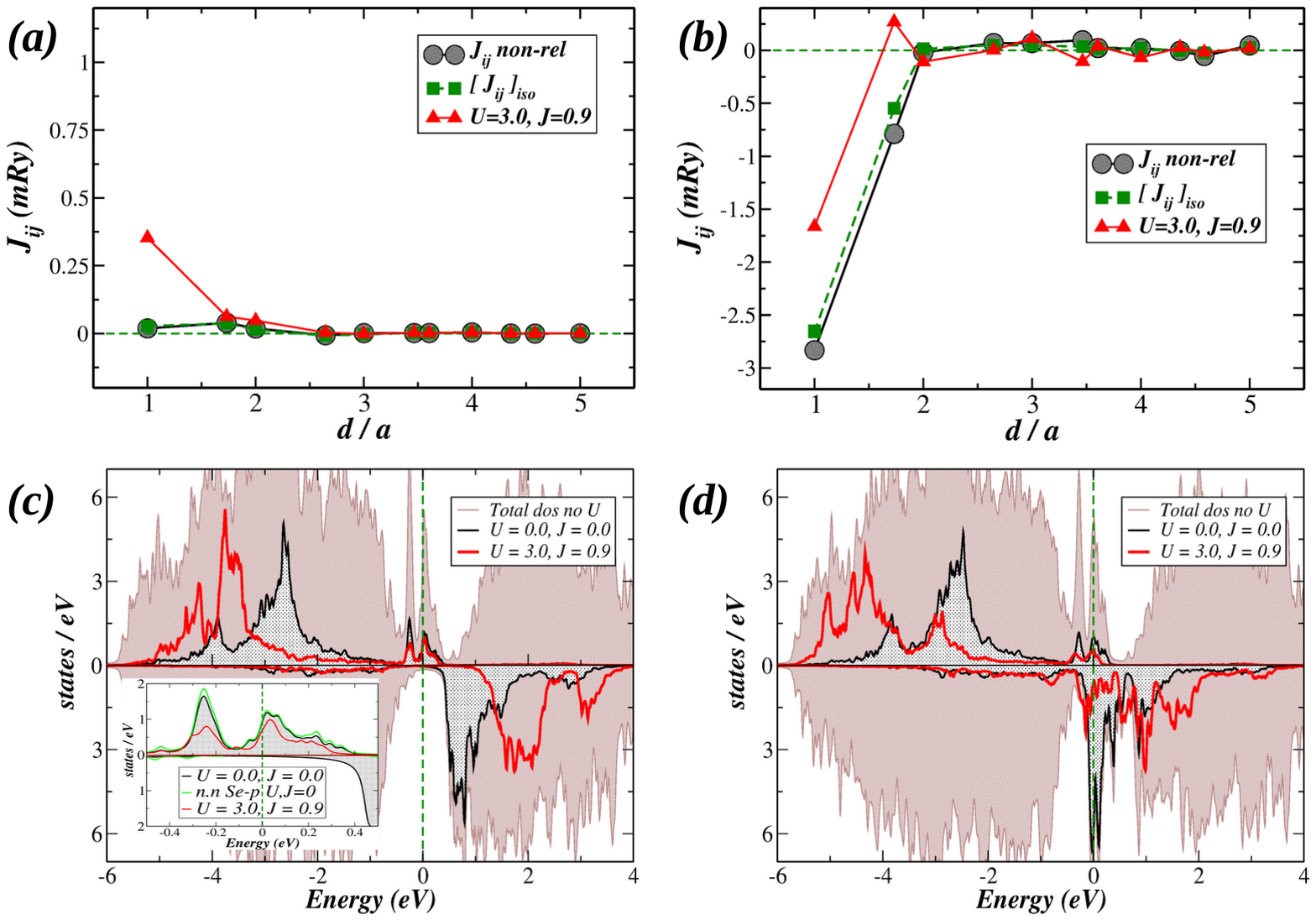}
\caption{TOP panel: Comparison between the calculated non-relativistic (with and without local correlation) magnetic exchange parameters and isotropic part of the magnetic exchange tensor (without local correlation) in the presence of SOC for the (a) Mn and (b) Fe doped (100\%) cases. BOTTOM panel: Local pdos with and without local correlation effect, corresponding to the (c) Mn and (d) Fe $3d$ states respectively. Inset in (c)  shows the Mn and Se states around Fermi energy and the strong hybridization between Mn $d$ and Se $p$ states. The grey-shaded region is the total dos without local correlation. pdos = projected density of states.}
\label{Bulk-100per_Mn-Fe}
\end{figure}
\FloatBarrier

The electronic and magnetic properties of Mn and Fe-doped systems are different from those of V and Cr-doped systems. This shall be clear from our discussion below. We start with the Mn-doped case. The charge occupation data of Mn without any local correlation, as reported in Table~\ref{Table4} shows 4.28 and 0.58 $e$ in the majority and minority $3d$-states respectively. This along with a local magnetic moment of 3.77 $\mu_B$ suggests a $d^4$ electron configuration with Mn in the 3+ oxidation state similar to the V and Cr cases. However, the pdos corresponding to the Mn $3d$ orbitals as shown in Fig.~\ref{Bulk-100per_Mn-Fe}(c) (black lines) do not fully agree with a $d^4$ electron configuration. The major contribution in the spin-up channel comes well below the Fermi energy in the -5 to -1 eV energy window. These states show a large exchange splitting w.r.t the spin-dn states suggesting a $d^5$ electron configuration. The small contribution around the Fermi energy in the form of two small peaks cannot represent an empty and filled $e_g$ state to support a $d^4$ situation. The total dos shown as the grey shaded region however shows two peaks around Fermi energy which are characteristic of a localized electron-hole pair. Comparing the total dos and Mn $3d$ states shows that the major contribution to this electron-hole states comes from the host, mainly the Se $p$ states as can be seen in the inset of  Fig.~\ref{Bulk-100per_Mn-Fe}(c) (green lines). This suggests the presence of a hole on the Se atoms near the Mn atom and hybridization with the Mn $e_g$ states gives it some $e_g$ character. Hence, Mn here is actually Mn2+ with a $d^5$ electron configuration and a hole in the host. The hole gets generated due to the substitution of Bi3+ ions with Mn2+. The reduced charge of 4.28 $e$ in the spin-up channel and a reduced Mn moment of 3.77 $mu_B$ result from the hybridization with a hole state. This is different from what we saw in the V and Cr cases.  This finding aligns with experimental studies of Mn-doped Bi$_2$Se$_3$, where XAS measurements demonstrate that Mn favours a 2+ oxidation state and acts as a $p$-type carrier system due to the presence of a hole\cite{Ychoi_APL.101_2012}. Also, similar reports are there in some of the previous theoretical studies\cite{MFIslam_PRB.97_2018,Yli_JCP.140_2014}. This behaviour of the Mn dopant is not unexpected and is very similar to the case of Mn substituted Ga in GaAs\cite{Nalmeleh_PRB.128_1962,JSzczytko_PPRB.60_1999,YSasaki_JAP.91_2022}. The contribution of the Mn $3d$ ($e_g$) to the hole state comes from the $p-d$ hybridization suggesting a possible hopping of the hole from Se to neighbouring Mn atoms. This can lead to a hole-mediated exchange between the Mn atoms. To understand the exchange mechanisms we again focus on the symmetry-resolved components of the nearest neighbour exchange as presented in Table~\ref{Table4}. Mn with a $d^5$ electron configuration in the high spin state suppresses the $e_g - T_{2g}$ FM superexchange. Now we get a weak $e_g - T_{2g}$ AFM superexchange for the nearly 90$^o$ Mn-Se-Mn exchange path according to GKA rules. This is very similar to what we see in compounds like MnCl$_2$, and MnBr$_2$ having a similar exchange path and Mn2+ ion\cite{Wang_2022-GKA}. The small AFM $T_{2g}-T_{2g}$  superexchange also remains as was also there for V and Cr cases. The $e_g - e_g$ interaction now becomes FM with an interaction strength comparable to the AFM $e_g - T_{2g}$ superexchange. This competition between FM $e_g - e_g$ and  AFM $T_{2g} - T_{2g}$, $e_g - T_{2g}$ interactions that almost cancel each other, results in a non-interacting situation as shown in Fig.~\ref{Bulk-100per_Mn-Fe}(a). Interestingly, the net n.n exchange even becomes weaker than the second neighbour exchange strength. A $d^5$ electron configuration is also expected to give a weak $e_g - e_g$ FM superexchange\cite{Wang_2022-GKA}, which was not present in the V and Cr cases. But in this case, the $e_g - e_g$  interaction is much more complex due to the presence of a hole in the system. Depending on the properties of the hole, two types of hole-mediated exchange can occur as explained in the Mn-doped Bi$_2$Te$_3$ theoretical findings\cite{Yli_JCP.140_2014}. A weakly bound hole leads to an RKKY-type long-range interaction, whereas a strongly bound hole is expected to result in an FM double exchange mechanism. To understand the nature of the $e_g - e_g$ component, we again performed an LDA+$U$ calculation similar to the V and Cr cases. This was done with the expectation that the hole state around Fermi energy would get perturbed and the corresponding changes in the exchange interaction would give us a clue in understanding its nature. Surprisingly, we find that the $e-h$ bound state near the Fermi level hardly gets affected,  as illustrated by the red lines in Fig.~\ref{Bulk-100per_Mn-Fe}(c). This confirms our claim that this state possesses more Se $p$-character and hence remains unaffected by local correlation applied on the Mn $3d$ states. As a result the hole-mediated FM $e_g - e_g$ interaction also remains unaffected (see Table~\ref{Table4}). The exchange splitting between spin-up and spin-dn states however increases due to the increased local correlation eliminating the AFM $e_g - T_{2g}$ and  $T_{2g}-T_{2g}$ superexchange interactions. Consequently, the total exchange interaction $J_{ij}$ now gets dominated by the FM $e_g - e_g$ component.  $J_{ij}$ as a function of distance, as illustrated in Fig.~\ref{Bulk-100per_Mn-Fe}(a) by the red lines shows a strongly decaying behaviour instead of a long-range oscillatory trend. This indicates that in our case the $e_g - e_g$ component comes from a short-range double exchange mechanism due to the strongly bound nature of the hole. This is expected in the case of Bi$_2$Se$_3$ as a host, which supports more localised properties compared to Bi$_2$Te$_3$ due to the smaller size of the Se $p$ orbitals\cite{Yli_JCP.140_2014, Jkim_PRB.96_2017}. Finally, we have also checked the effect of SOC on the exchange in this case without any local correlation effect. Similar to V and Cr cases, the effect is negligible as can be seen from the comparison given in Fig.~\ref{Bulk-100per_Mn-Fe}(a).

\begin{figure}[h]
\includegraphics[scale=0.55]{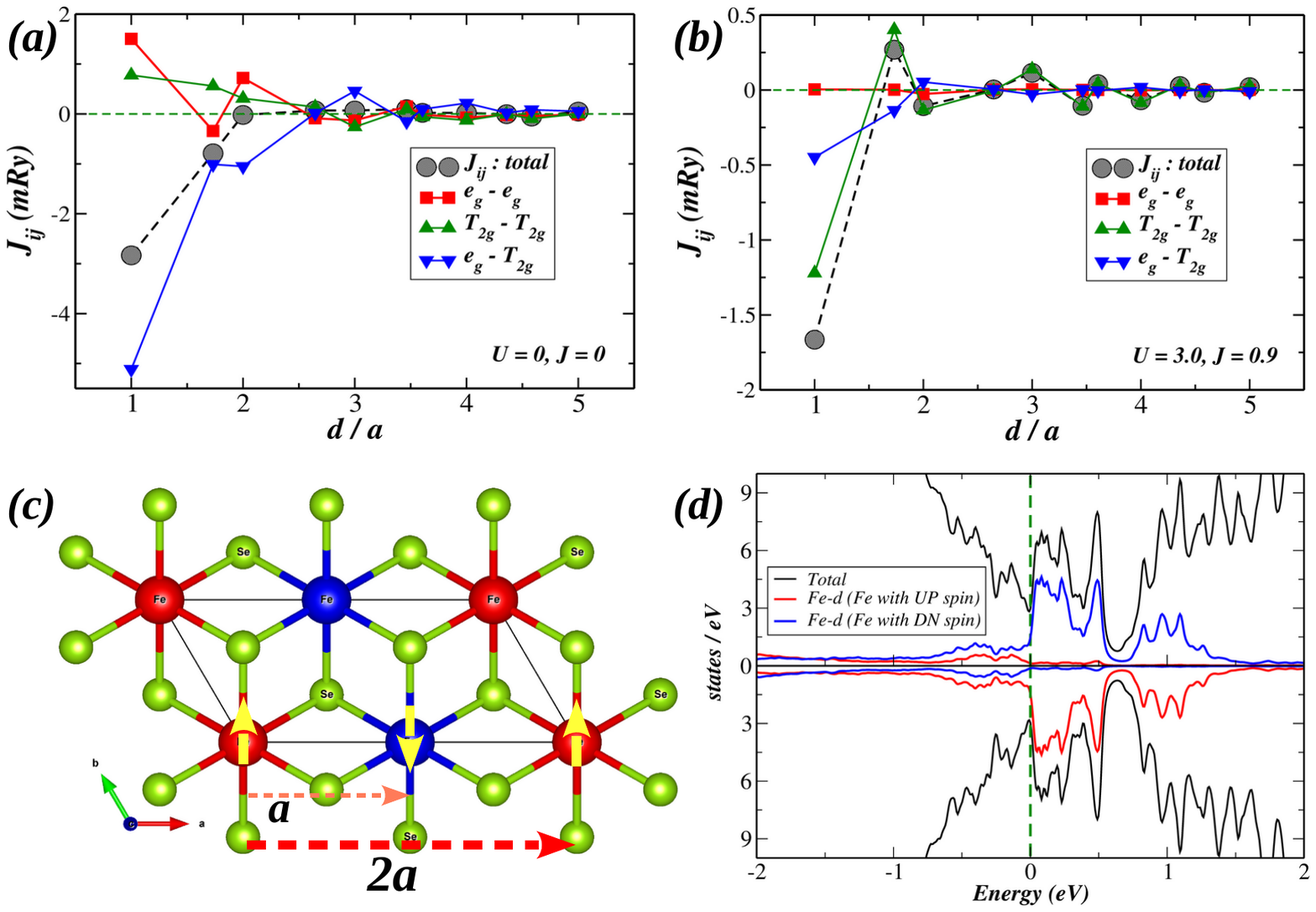}
\caption{Orbital resolved exchange components as a function of distance for the 100\% Fe doped case (a) without and (b) with local correlation effects on the Fe $3d$ states. (c) A simple AFM ordering of the Fe atoms considering a unit cell doubling along the $a$-axis. Fe atoms are shown with red and blue spheres. Se atoms are shown with green spheres. (d) The total and local pdos of the Fe atoms from this AFM solution.}
\label{Fe-Bulk-100per_OR}
\end{figure}

Going from Mn to Fe adds one more $d$-electron in the system. Without any local correlation effect on the Fe $3d$ states, the charge distribution in both the spin channels and local magnetic moment as given in Table~\ref{Table4} suggests a $d^5$-up, $d^1$-dn electron configuration. The corresponding local pdos are shown by the black lines in Fig.~\ref{Bulk-100per_Mn-Fe}(d). The structure of the spin-up channel is very similar to the Mn case with the presence of an electron-hole state around the Fermi energy suggesting a $d^5$ electron configuration. But here in the case of Fe,  these states look a bit more delocalized with a larger contribution at the Fermi level. This happens possibly because of the smaller size of the Fe atoms and a stronger $p-d$ hybridization with the Se atoms. Along with this, the down-spin channel also contributes at the Fermi level which mainly comes from the Fe $d$ states again supporting the $d^1$-dn electron configuration. These data show that the Fe-doped system behaves similarly to the Mn-doped system in terms of oxidation state. However, the presence of this extra $d$ electron in the down spin channel and the corresponding electronic structure as discussed above drastically changes the nature of the magnetic exchange interaction. As shown in Fig.~\ref{Bulk-100per_Mn-Fe}(b) (black spheres), we now get a strong AFM n.n exchange followed by an AFM second neighbour exchange of significant strength. The AFM exchange and a second neighbour exchange of such strength were missing in the previous three systems with V, Cr, and Mn. These AFM interactions were also found in a previous theoretical study of Fe doped Bi$_2$Se$_3$, Bi$_2$Te$_3$, and Sb$_2$Te$_3$\cite{verginory_PhysRevB}. Besides this, we also see a distant oscillatory behaviour which becomes more prominent in the case of LDA+$U$ calculations as shown by the red lines in Fig.~\ref{Bulk-100per_Mn-Fe}(b). This is a signature of the RKKY-type exchange mechanism. To understand the origin of the AFM exchange we again focus on the orbital resolved components of the n.n exchange as reported in Table~\ref{Table4}. At the same time to understand the long-range oscillatory behaviour, we also check the distance dependence of these orbital resolved components as presented in Fig.~\ref{Fe-Bulk-100per_OR}. Before discussing the orbital components we would like to point out that in this case, the SOC has a larger impact on the first and second neighbour exchange parameters compared to the previous systems. This could happen for several reasons for example structural distortions, the presence of larger anisotropic exchange terms etc and needs further investigations which are not a part of this study.

\FloatBarrier

The $e_g - e_g$ component in this case also gives a hole-mediated FM n.n exchange similar to the Mn-doped case. But in this case, the strength is larger compared to the Mn-doped case. This component  also shows a long-ranged RKKY-type oscillatory trend as shown in Fig.~\ref{Fe-Bulk-100per_OR}(a) (red squares) suggesting a weakly bound hole similar to what was suggested in the case of Mn-doped Bi$_2$Te$_3$\cite{Yli_JCP.140_2014}. Hence, the hole, in this case, could result in both short-range FM double exchange and long-ranged RKKY type exchange. And due to the more delocalized nature of the hole, the n.n $e_g - e_g$ component is larger compared to the Mn-doped case. Next, we focus on the $T_{2g} - T_{2g}$ component (green up-triangles). It has a n.n FM interaction that decays with distance followed by a weak oscillatory behaviour. The short-ranged FM interactions occur as a result of a double exchange mechanism due to the finite contribution at the Fermi level coming from the singly occupied $T_{2g}$-dn states (Fig.~\ref{Bulk-100per_Mn-Fe}(d)). The double exchange dominates over RKKY due to the strongly localized impurity-like nature of this state. Finally, we discuss the $e_g - T_{2g}$ component (blue dn-triangles) that gives the major and dominating AFM interactions between two Fe atoms. The total density of states (Fig.~\ref{Bulk-100per_Mn-Fe}(d)) shows a large contribution to the Fermi energy in both the majority and minority spin channels. The states in the majority spin channel are hole-like states with significant $e_g$ characters as discussed before. On the other hand, the states in the minority spin channel are electrons coming from the partially filled $T_{2g}$-dn band. Both these states are localized in nature. Now, an AFM order between two Fe atoms will make it possible for the holes and electrons to hop from the majority spin channel of one Fe to the minority spin channel of the other Fe atom and vice versa. This hopping will happen via the Se atoms and will lead to hybridization between two different spin channels. This will make the system more metallic with more delocalized charge carriers promoting kinetic energy gain. This is more like an antiferromagnetic double-exchange mechanism that becomes possible in this system because of the 2+ oxidation state of the Fe atom and a $d^6$ electron configuration. To check if what we say is correct,  we have performed an electronic structure calculation with a simple AFM state by doubling the unit cell along the $a$-axis and putting AFM spin arrangement between the two Fe atoms. This is shown schematically in Fig.~\ref{Fe-Bulk-100per_OR}(c). The total and local pdos of the Fe atoms from this AFM calculation are shown in Fig.~\ref{Fe-Bulk-100per_OR}(d). The solution comes out to be lower in energy compared to the FM solution we discussed so far. Also as expected we get a fully metallic solution with a more delocalized state around Fermi energy compared to the FM solution (Fig.~\ref{Bulk-100per_Mn-Fe}(d)). One shall note in this case that the Fe atoms here are forming a triangular lattice and hence the correct AFM ground state is going to be much more complex than the one we have considered here.  To further verify that our interpretations of the exchange mechanisms present in this Fe-doped system are correct, we have again performed an LDA+$U$ calculation to manipulate the electronic structure around Fermi energy and then observe the induced changes in the exchange interaction. The distance dependence of the exchange components with the local correlation effect is shown in Fig.~\ref{Fe-Bulk-100per_OR}(b) and the local pdos corresponding to the Fe $d$-state is shown in red in Fig.~\ref{Bulk-100per_Mn-Fe}(d). With an increased local correlation the exchange splitting of the Fe $d$-state increases and the hole-like state at the Fermi energy almost gets killed. This is different compared to the Mn case where it was almost unaffected. This happens as a result of larger $p-d$ hybridization and more $3d$ character of the hole state compared to the Mn case. The vanishing hole state is no longer able to give rise to the hole-mediated double exchange and as a result the $e_g - e_g$ component of the exchange interaction totally dissipates as can be seen in Fig.~\ref{Fe-Bulk-100per_OR}(b). The local correlation also has a significant effect on the $T_{2g}$ state in the minority spin channel. Interestingly, it becomes more delocalized around Fermi energy and reduces its contribution to the Fermi level.  A reduced contribution at the Fermi level from both the $e_g$ like hole state in the majority spin channel and $T_{2g}$ state in the minority spin channel reduces the strength of the AFM double exchange type $e_g - T_{2g}$ component. It no longer remains the major source of AFM n.n exchange. Now the major contribution comes from the $T_{2g} - T_{2g}$ component that also shows a prominent long-ranged oscillatory behaviour. Though the $T_{2g} - T_{2g}$ shows an n.n AFM exchange, the second neighbour exchange is FM suggesting a dominating RKKY mechanism. This happens due to the delocalization of the $T_{2g}$ state with a correlation effect. Now, the net exchange of the system also gets governed by the $T_{2g} - T_{2g}$ component with long-ranged RKKY behaviour. Interestingly, a very similar trend of the $T_{2g} - T_{2g}$ component was reported in the bulk bcc-Fe\cite{YOKvashnin_PRL_116.2016}. These results suggest the presence of competing exchange interactions in the Fe system and the possibility of tuning the magnetic order of the system from AFM to FM with an application of local perturbations. Though a higher doping concentration of Fe doping is not suitable for QAHE due to the AFM  ground state, with a lower doping concentration the presence of this RKKY-type interaction can play a significant role in the stability of an FM ordering which needs further investigations.  We will try to shed some light on this in the next section where we consider a lower 50\% doping concentration.

\subsubsection{The 50\% doped case}

\begin{figure}[h]
\includegraphics[scale=0.55]{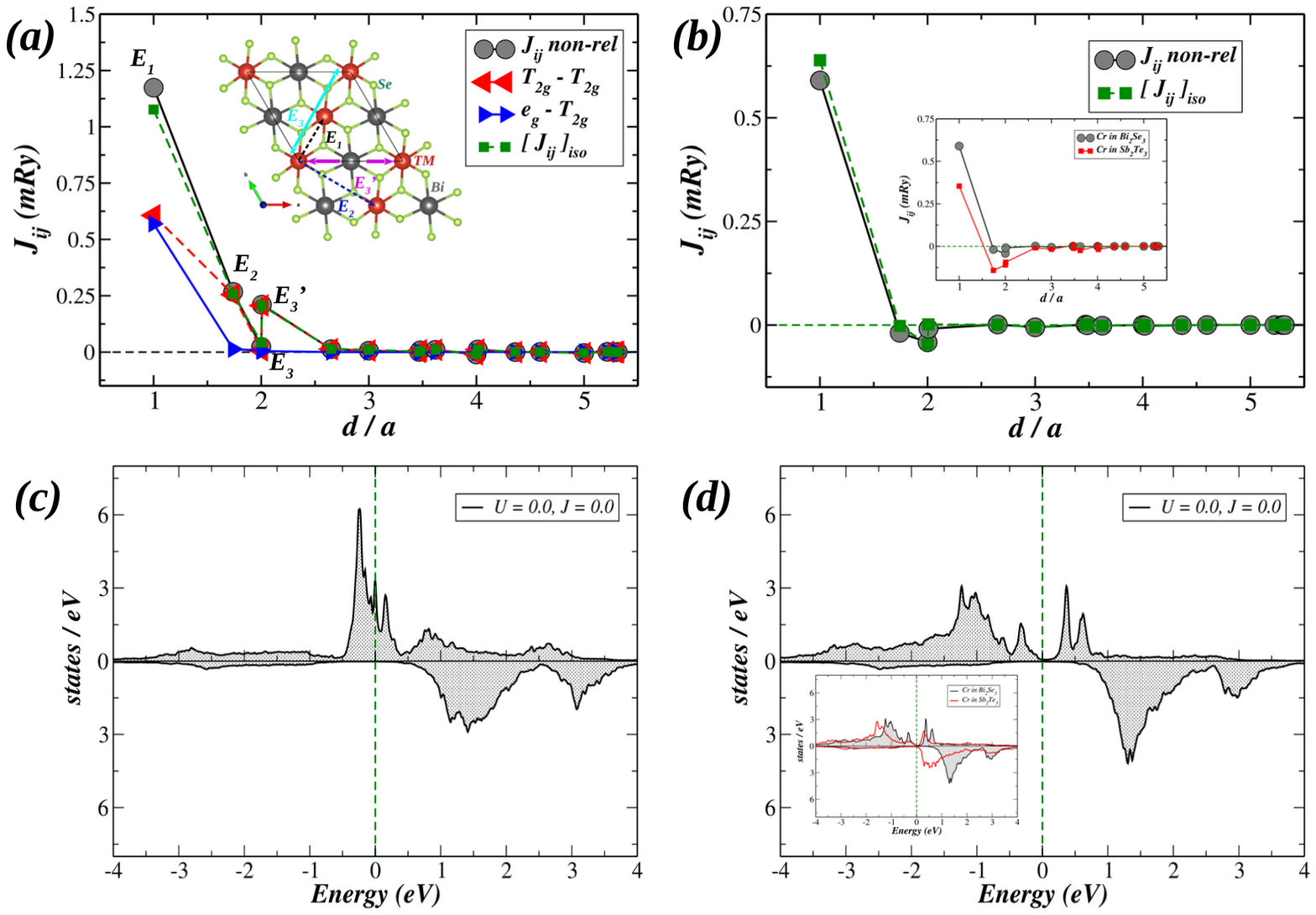}
\caption{TOP panel: The calculated non-relativistic magnetic exchange parameters and isotropic part of the magnetic exchange tensor in the presence of SOC for the (a) V and (b) Cr doped (50\%) cases. The inset of (a) shows the exchange paths for a 50\% doping that has been discussed in the main text. The inset of (b) shows the calculated exchange parameters for Cr doped Sb$_2$Te$_3$. BOTTOM panel: Local pdos without any local correlation corresponding to the $3d$ states of (c) V and (d) Cr. The inset of (d) shows the calculated local pdos for Cr doped Sb$_2$Te$_3$. pdos = projected density of states.}
\label{Bulk-50per_V-Cr}
\end{figure}

Our main results on the different type of exchange mechanisms that results depending on the type of magnetic dopant, and the harmony or competition between them have been discussed in the previous section. In this section, we focus on the effect of doping concentration on these exchange mechanisms. To this end, we now consider the 50\% doping case as shown schematically in Fig.~\ref{Struc_dop-tech-crop}(b). In this case, the average TM-TM distance increases, but we want to focus more on the local structural arrangements of the dopant atoms. If we compare with the 100\% doping case, we can see that in this case also we have the same n.n exchange path (TM-Se-TM / $E_1$). Hence, we expect that the n.n exchange in this case is also going to be more or less the same as the 100\% doping case. The difference occurs in the third neighbour exchange. In the 100\% doping case we only had one type of third neighbor exchange path, TM-Se-TM-Se-TM / $E_3$. But now for the 50\% doping case, we have two different types of paths, the first one and a new one, TM-Se-Bi-Se-TM / $E_3'$. The second neighbour exchange path $E_2$ in the 50\% doped case also becomes different with the presence of Bi atoms nearby. These paths are schematically shown in the inset of Fig.~\ref{Bulk-50per_V-Cr}(a).  Our aim in this case is to check how the magnetic exchange along $E_3'$ differs from the one along $E_3$. Especially, along $E_3'$ we have the direct presence of the Bi atom that could produce a significant SOC effect. This will allow us to directly check the importance of the van Vleck exchange mechanism which was not possible in the 100\% doping case. For this 50\% doping case, the structures were fully optimized in VASP as before. The details of the structural changes are not given here but have been discussed in section III of the SM. The most visible structural change is the increase in the unit cell volume compared to the 100\% doped case which is expected as a result of more Bi content. Also in this case, the unit cell volume and lattice parameters come out to be almost the same for each dopant type (V, Cr, Mn, and Fe). This is something different from the 100\% doping case but is also expected because, for lower doping like this, the volumetric properties shall depend more on the host and not on the dopants. For electronic structure calculation in this case, we have done only LDA calculations without any local correlation effects.

\FloatBarrier

We start with the V and Cr systems. The calculated exchange parameters and local pdos corresponding to the $3d$ states are shown in Fig.~\ref{Bulk-50per_V-Cr}. For both V and Cr, we again see a 3+ oxidation state with a $d^2$ and $d^3$ electron configuration respectively. As a result, we again get a finite contribution from the $T_{2g}$-up states at the Fermi level in the V case as shown in Fig.~\ref{Bulk-50per_V-Cr}(c). For Cr, we get the same insulating solution as before as shown in Fig.~\ref{Bulk-50per_V-Cr}(d). The magnetic exchange as a function of distance for the V case is shown in Fig.~\ref{Bulk-50per_V-Cr}(a). In this case, we see a significant difference compared to the 100\% doped case shown in Fig.~\ref{Bulk-100per_V-Cr}(a). The n.n FM interaction is very similar to the 100\% doped case  and comes from the  combination of $T_{2g} - T_{2g}$ double exchange and $e_g - T_{2g}$ superexchange as before. But the second neighbour and the third neighbour interaction along exchange path $E_3'$ shows a larger FM contribution. The presence of the Bi atoms along these paths initially points to the van Vleck mechanism as the reason. But we see that in the presence of SOC (green squares), they remain almost unaffected suggesting a different mechanism. If we now focus on the Cr case as shown in Fig.~\ref{Bulk-100per_V-Cr}(b), we do not see these FM second and third neighbour interactions and the magnetic exchange as a function of distance is very similar to the 100\% doped case (Fig.~\ref{Bulk-50per_V-Cr}(b)). However,  the exchange paths in both V and Cr systems are the same. This suggests the role of the partially filled $T_{2g}$-up states in the V system for this enhanced second and third neighbour FM interaction strength. Analysis of the orbital components further shows that these interactions come only from the $T_{2g} - T_{2g}$ component as shown by the left triangles in Fig.~\ref{Bulk-50per_V-Cr}(a). Hence, in this case, the origin is not van Vleck but the short-range double exchange mechanism. In the Cr case, it is not present as before due to the insulating nature and we only see the n.n FM superexchange. The second and third neighbour interactions in the Cr case are weakly AFM instead of FM. This is however not something new and was previously reported for Cr-doped Bi$_2$Se$_3$\cite{Jkim_PRB.96_2017} similar to our results. The second and third neighbour interactions in the Cr case occur from the polarized $p$-orbital network\cite{Jkim_PRB.96_2017, Jkim_PRB.97_2018} of the host without getting affected by SOC. The origin of the AFM nature was linked to the properties of the Se $p$ orbitals in Bi$_2$Se$_3$\cite{Jkim_PRB.96_2017} host and was found to the FM when Cr was doped in Bi$_2$Te$_3$ and Sb$_2$Te$_3$. To check this we did a model calculation with our Cr system by replacing Bi$_2$Se$_3$ with Sb$_2$Te$_3$ keeping the structural degrees of freedom fixed. The electronic structure and exchange parameters in a comparative manner are shown in the inset of Fig.~\ref{Bulk-50per_V-Cr} (d) and (b) respectively.  We see that the system remains insulating preventing the carrier-mediated mechanisms (double exchange, RKKY) from occurring. The red squares show the magnetic exchange data for the Sb$_2$Te$_3$ case. The second neighbour and third neighbour exchange parameters however become more AFM in nature instead of FM. This could happen because the structure was not fully optimized after replacing Bi$_2$Se$_3$ with Sb$_2$Te$_3$. However, these changes confirm the active role of the polarized $p$-orbital network of the host in mediating distant exchange interactions which is much weaker compared to the carrier-driven mechanisms. Coming back to our real system, though we see no effect of SOC on the exchange along $E_2$, $E_3$, and $E_3'$, the n.n exchange for the 50\% doped case does show greater changes with SOC compared to the 100\% doped case. But we shall keep in mind that the n.n exchange mechanisms mediate through the Se atoms which are now also bonded with the Bi atoms. Hence, with SOC the Se $p$ states probably get more affected leading to these larger changes. But looking at the above results we could conclude that the van Vleck mechanism does not have any significant role in this system in mediating a distant FM exchange between the magnetic atoms. Next, we are going to discuss the Mn and Fe case.

\begin{figure}[h]
\includegraphics[scale=0.55]{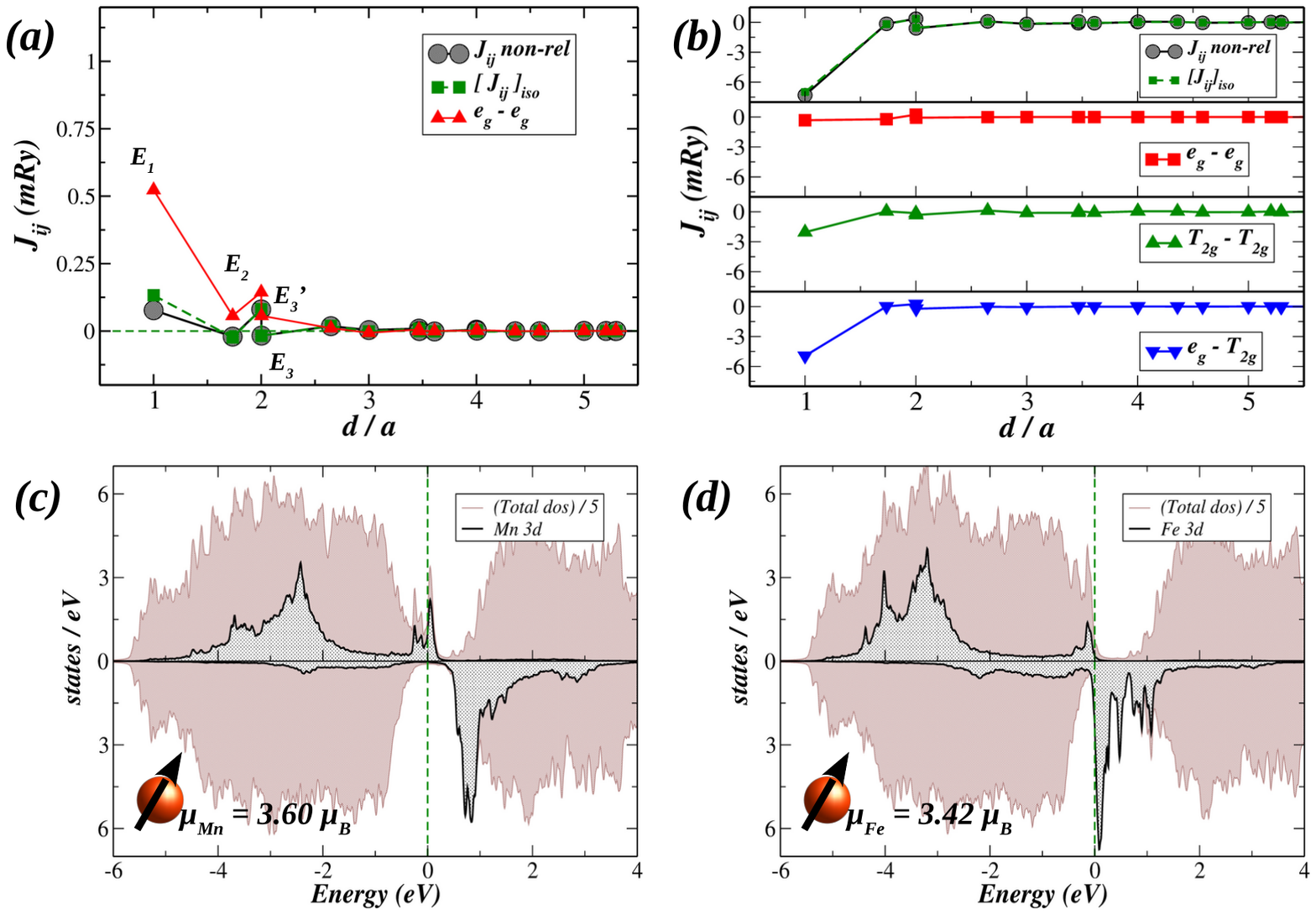}
\caption{TOP panel: The calculated non-relativistic magnetic exchange parameters and isotropic part of the magnetic exchange tensor in the presence of SOC, as a function of distance for the (a) Mn and (b) Fe doped (50\%) cases. For the Fe case, the total and orbital decomposed components are shown together in separate panels for the same energy window. BOTTOM panel: Local pdos without any local correlation corresponding to the $3d$ states of (c) Mn and (d) Fe. The local magnetic moments are also indicated. pdos = projected density of states.}
\label{Bulk-50per_Mn-Fe}
\end{figure}
\FloatBarrier

For the Mn-doped case, we again see a local pdos and magnetic moment similar to the 100\% doped case as shown in Fig.~\ref{Bulk-50per_Mn-Fe}(c). This says that Mn is in the 2+ oxidation state. But in this case, the hole state shows a more delocalized nature with a larger contribution to the Fermi energy. This turns the hole-mediated $e_g - e_g$ double exchange component to dominate over the n.n AFM superexchange components making the net n.n exchange weakly FM as shown in Fig.~\ref{Bulk-50per_Mn-Fe}(a). The $e_g - e_g$ component has been shown with red up-triangles for a better understanding and we can see a contribution to the third neighbour exchange along the exchange path $E_3'$. But the overall exchange interactions are as weak as before and things are not so interesting. We also see negligible effect from SOC as before. The finite contribution at Fermi energy and the presence of a double exchange mechanism prevents us from checking the role of the $p$-orbital mediated exchange for the Mn case. Things become interesting for the Fe system. As shown in Fig.~\ref{Bulk-50per_Mn-Fe}(d), the local magnetic moment and pdos are very different compared to the 100\% doped case. Now, Fe looks more like Fe3+ with a $d^5$ electron configuration and the system is very close to an insulating state. The hole state in the majority spin channel has almost disappeared. As a result, the hole-mediated $e_g - e_g$ double exchange is no longer active as shown in the second panel of Fig.~\ref{Bulk-50per_Mn-Fe}(b). The contribution from the $T_{2g}$-dn state at the Fermi level also reduces significantly, eliminating the short-ranged FM $T_{2g} - T_{2g}$ double exchange that we saw in the 100\% case. This is shown in the third panel of Fig.~\ref{Bulk-50per_Mn-Fe}(b).  Though a small finite contribution at the Fermi level from the $T_{2g}$ state leads to a very weak long-ranged oscillation, the stronger AFM n.n value cannot come from the RKKY mechanism. The almost insulating state suggests a n.n $T_{2g} - T_{2g}$ direct exchange or superexchange mechanism. Still, due to the small finite contribution of $e_g$-up and $T_{2g}$-dn at the Fermi level, the AFM double exchange type $e_g - T_{2g}$ component remains as before giving rise to the dominant AFM n.n exchange. This is shown in the last panel.

\begin{figure}[h]
\includegraphics[scale=0.55]{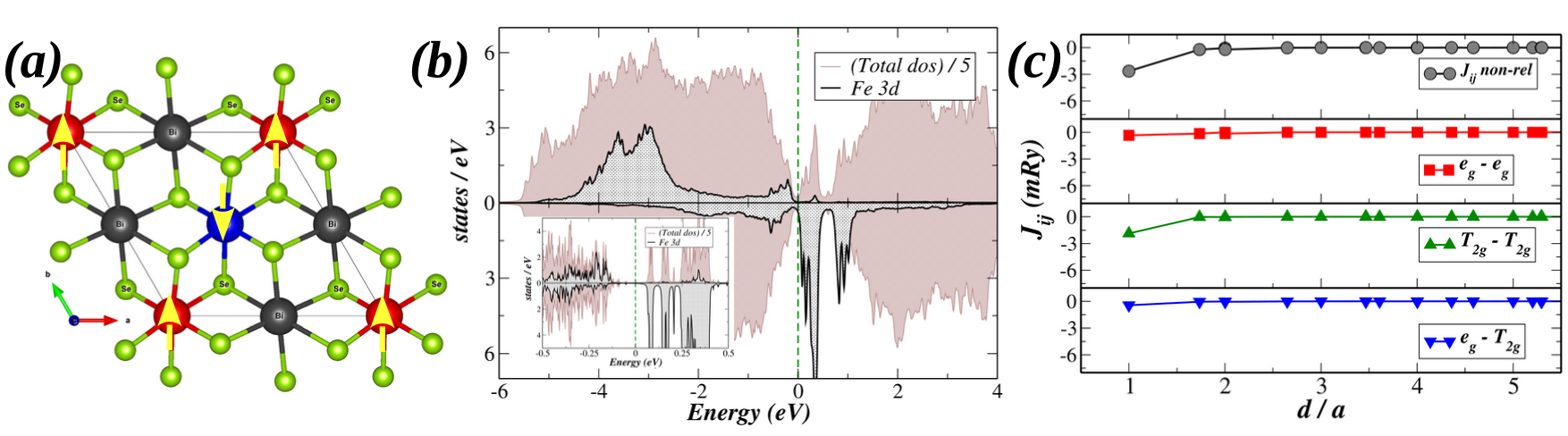}
\caption{(a) The AFM ordering of the Fe atoms in the 50\% doping case. (b) Total and local pdos corresponding to the Fe $3d$ states without any local correlation for this AFM ordering. The inset shows the states around Fermi energy calculated using smaller smearing illustrating the insulating state of the system. (c) Distance dependence of the total and orbital resolved components of the exchange interactions showing the presence of only n.n $T_{2g} - T_{2g}$ AFM direct exchange. pdos = projected density of states.}
\label{Bulk-50per_Fe-AFM-crop}
\end{figure}
\FloatBarrier

To further check and verify these interactions, we again did a calculation with AFM ordering of the Fe atoms as shown in Fig.~\ref{Bulk-50per_Fe-AFM-crop}(a). We shall note that this AFM order is the correct order for the 50\% doping case as there is no lattice frustration. In the 100\% doping case, the AFM order we considered was a simple one but it was not the actual ground state AFM order. The total energy of the AFM solution comes out to be lower than the FM solution as expected. The calculated total and Fe-$3d$ projected local dos are shown in Fig.~\ref{Bulk-50per_Fe-AFM-crop}(b). The total and orbital decomposed exchange parameters as a function of distance are shown in Fig.~\ref{Bulk-50per_Fe-AFM-crop}(c). As can be seen from the dos the system looks insulating with no contribution from the Fe-$3d$ states in the majority spin channel. Some contributions from the $T_{2g}$-dn state that we see results from a technical issue due to the higher value of smearing that we use while printing the spectral properties. To confirm, we have also calculated the dos with smaller smearing and the states around Fermi energy are shown in the inset of Fig.~\ref{Bulk-50per_Fe-AFM-crop}(b) illustrating the insulating state of the system. Now if we focus on the exchange we see additional changes due to this AFM insulating state. The $e_g - e_g$ component is again not contributing as expected. But now we see that the $e_g-T_{2g}$ n.n AFM exchange gets suppressed. This is because the system is insulating and the AFM double exchange is no longer possible. Though the $T_{2g} - T_{2g}$ nearest neighbour AFM interaction remains almost unaffected, the weak long-ranged oscillatory part that was there for the FM case vanishes. The oscillatory part is understood as the contribution from the $T_{2g}$ states at Fermi energy is no longer there. Hence, the nearest neighbour AFM exchange that remains almost unaffected going from FM to AFM order cannot be RKKY is now confirmed. It could be either a superexchange or a direct exchange as we guessed before. In our structure, the Se atoms do not make a Fe-Se-Fe 180$^o$ exchange path to produce a strong AFM n.n superexchange. So, this is a direct exchange between the neighbouring Fe atoms leading to the AFM order. This n.n AFM direct exchange could also be present in the 100\% doped case but gets outweighed by the carrier-mediated double or RKKY mechanisms. Hence, the 50\% doped data suggests that the metallic solution that we got in the 100\% doped case (which was even more metallic with the AFM order) is a consequence of the closely spaced Fe atoms on a triangular lattice. But we have to consider the fact that we never considered the correct AFM order in the 100\% doped case. The correct order is supposed to be some non-collinear order and with that, the system could become insulating, killing the double and RKKY exchange. In that case, we might get an AFM state due to the n.n direct exchange even for the 100\% doping. But this needs further investigation. Hence we can conclude that, for Fe, things are complex and the exchange interactions are expected to depend strongly on the doping concentration as well as local correlation effects as we saw in the 100\% doping case. Our data shows that with lower doping concentrations when we have a lower probability of closely spaced n.n Fe atoms, there will be no dominant AFM interaction. The system could be FM due to other mechanisms like the one that gets mediated by the polarized $p$-orbital network. This needs further investigation and is not a part of this work.

\section{Conclusion}

Taking advantage of the local octahedral symmetry we have calculated the orbital decomposed exchange interactions between the doped TM atoms inside a quintuple layer of Bi$_2$Se$_3$. Correlating these parameters with the calculated electronic structure for V, Cr, Mn, and Fe doped Bi$_2$Se$_3$, we have identified efficiently and directly, the various types of exchange mechanisms that have been discussed in the literature so far. Considering the local correlation parameter as a tool to manipulate the local electronic structure and then measuring the changes in the exchange parameters we are able to show the sensitivity of these exchange mechanisms on the electronic structure and how their nature could change from one type to another. We find that Cr is the only system that shows a robust insulating state with ferromagnetic exchange that remains unaffected by local correlation and doping concentration. For such a system, exchange interaction other than the n.n interaction gets mediated via the polarized $p$-orbital network of the host.  The rest of the systems, V, Mn, and Fe, as a result of their electronic configuration and local octahedral environment always have a possibility of finite carrier density at the Fermi energy. Depending on the type of carrier (electron/hole) and their localized/delocalized nature, a short-ranged double exchange / long-ranged RKKY mechanism could occur. For V, we get a strong electron-mediated double exchange, whereas for Mn, we get a weaker hole-mediated double exchange that could control the distant neighbour interactions. Unlike others, for Fe, we observed strong AFM exchange interaction. The basis of the exchange mechanism is complex and strongly depends on the interplay between local correlation and doping concentration.  Hence, for Fe, to get the correct magnetic ground state at any doping concentration shall require realistic modelling of the doped system, for example via  Special Quasi-random Structures (SQS) generation. And the use of correct values of the local correlation parameters that could be calculated for example using the linear response approach.

\bibliographystyle{apsrev}

\bibliography{references}

\begin{thebibliography}{80}
\expandafter\ifx\csname natexlab\endcsname\relax\def\natexlab#1{#1}\fi
\expandafter\ifx\csname bibnamefont\endcsname\relax
  \def\bibnamefont#1{#1}\fi
\expandafter\ifx\csname bibfnamefont\endcsname\relax
  \def\bibfnamefont#1{#1}\fi
\expandafter\ifx\csname citenamefont\endcsname\relax
  \def\citenamefont#1{#1}\fi
\expandafter\ifx\csname url\endcsname\relax
  \def\url#1{\texttt{#1}}\fi
\expandafter\ifx\csname urlprefix\endcsname\relax\def\urlprefix{URL }\fi
\providecommand{\bibinfo}[2]{#2}
\providecommand{\eprint}[2][]{\url{#2}}

\bibitem[{\citenamefont{Klitzing et~al.}(1980)\citenamefont{Klitzing, Dorda, and Pepper}}]{klitzing_PhysRevLett.45.494_1980}
\bibinfo{author}{\bibfnamefont{K.~v.} \bibnamefont{Klitzing}}, \bibinfo{author}{\bibfnamefont{G.}~\bibnamefont{Dorda}}, \bibnamefont{and} \bibinfo{author}{\bibfnamefont{M.}~\bibnamefont{Pepper}}, \bibinfo{journal}{Phys. Rev. Lett.} \textbf{\bibinfo{volume}{45}}, \bibinfo{pages}{494} (\bibinfo{year}{1980}), \urlprefix\url{https://link.aps.org/doi/10.1103/PhysRevLett.45.494}.

\bibitem[{\citenamefont{Liu et~al.}(2008)\citenamefont{Liu, Qi, Dai, Fang, and Zhang}}]{liu_PhysRevLett.101.146802_2008}
\bibinfo{author}{\bibfnamefont{C.-X.} \bibnamefont{Liu}}, \bibinfo{author}{\bibfnamefont{X.-L.} \bibnamefont{Qi}}, \bibinfo{author}{\bibfnamefont{X.}~\bibnamefont{Dai}}, \bibinfo{author}{\bibfnamefont{Z.}~\bibnamefont{Fang}}, \bibnamefont{and} \bibinfo{author}{\bibfnamefont{S.-C.} \bibnamefont{Zhang}}, \bibinfo{journal}{Phys. Rev. Lett.} \textbf{\bibinfo{volume}{101}}, \bibinfo{pages}{146802} (\bibinfo{year}{2008}), \urlprefix\url{https://link.aps.org/doi/10.1103/PhysRevLett.101.146802}.

\bibitem[{\citenamefont{Wu et~al.}(2014)\citenamefont{Wu, Liu, and Liu}}]{wu_PhysRevLett.113.136403_2014}
\bibinfo{author}{\bibfnamefont{J.}~\bibnamefont{Wu}}, \bibinfo{author}{\bibfnamefont{J.}~\bibnamefont{Liu}}, \bibnamefont{and} \bibinfo{author}{\bibfnamefont{X.-J.} \bibnamefont{Liu}}, \bibinfo{journal}{Phys. Rev. Lett.} \textbf{\bibinfo{volume}{113}}, \bibinfo{pages}{136403} (\bibinfo{year}{2014}), \urlprefix\url{https://link.aps.org/doi/10.1103/PhysRevLett.113.136403}.

\bibitem[{\citenamefont{Haldane}(1988)}]{haldane_PhysRevLett.61.2015_1988}
\bibinfo{author}{\bibfnamefont{F.~D.~M.} \bibnamefont{Haldane}}, \bibinfo{journal}{Phys. Rev. Lett.} \textbf{\bibinfo{volume}{61}}, \bibinfo{pages}{2015} (\bibinfo{year}{1988}), \urlprefix\url{https://link.aps.org/doi/10.1103/PhysRevLett.61.2015}.

\bibitem[{\citenamefont{Yu et~al.}(2010{\natexlab{a}})\citenamefont{Yu, Zhang, Zhang, Zhang, Dai, and Fang}}]{rui_science.329.5987_2010}
\bibinfo{author}{\bibfnamefont{R.}~\bibnamefont{Yu}}, \bibinfo{author}{\bibfnamefont{W.}~\bibnamefont{Zhang}}, \bibinfo{author}{\bibfnamefont{H.-J.} \bibnamefont{Zhang}}, \bibinfo{author}{\bibfnamefont{S.-C.} \bibnamefont{Zhang}}, \bibinfo{author}{\bibfnamefont{X.}~\bibnamefont{Dai}}, \bibnamefont{and} \bibinfo{author}{\bibfnamefont{Z.}~\bibnamefont{Fang}}, \bibinfo{journal}{Science} \textbf{\bibinfo{volume}{329}}, \bibinfo{pages}{61} (\bibinfo{year}{2010}{\natexlab{a}}), \eprint{https://www.science.org/doi/pdf/10.1126/science.1187485}, \urlprefix\url{https://www.science.org/doi/abs/10.1126/science.1187485}.

\bibitem[{\citenamefont{Zhang et~al.}(2016)\citenamefont{Zhang, Hsu, and Liu}}]{zhang_PhysRevB.93.235315_2016}
\bibinfo{author}{\bibfnamefont{R.-X.} \bibnamefont{Zhang}}, \bibinfo{author}{\bibfnamefont{H.-C.} \bibnamefont{Hsu}}, \bibnamefont{and} \bibinfo{author}{\bibfnamefont{C.-X.} \bibnamefont{Liu}}, \bibinfo{journal}{Phys. Rev. B} \textbf{\bibinfo{volume}{93}}, \bibinfo{pages}{235315} (\bibinfo{year}{2016}), \urlprefix\url{https://link.aps.org/doi/10.1103/PhysRevB.93.235315}.

\bibitem[{\citenamefont{Zhang et~al.}(2009)\citenamefont{Zhang, Liu, Qi, Dai, Fang, and Zhang}}]{zhang_natphy_5.6_2009}
\bibinfo{author}{\bibfnamefont{H.}~\bibnamefont{Zhang}}, \bibinfo{author}{\bibfnamefont{C.-X.} \bibnamefont{Liu}}, \bibinfo{author}{\bibfnamefont{X.-L.} \bibnamefont{Qi}}, \bibinfo{author}{\bibfnamefont{X.}~\bibnamefont{Dai}}, \bibinfo{author}{\bibfnamefont{Z.}~\bibnamefont{Fang}}, \bibnamefont{and} \bibinfo{author}{\bibfnamefont{S.-C.} \bibnamefont{Zhang}}, \bibinfo{journal}{Nature physics} \textbf{\bibinfo{volume}{5}}, \bibinfo{pages}{438} (\bibinfo{year}{2009}).

\bibitem[{\citenamefont{Chang et~al.}(2013{\natexlab{a}})\citenamefont{Chang, Zhang, Feng, Shen, Zhang, Guo, Li, Ou, Wei, Wang et~al.}}]{cui_science.340.167_2013}
\bibinfo{author}{\bibfnamefont{C.-Z.} \bibnamefont{Chang}}, \bibinfo{author}{\bibfnamefont{J.}~\bibnamefont{Zhang}}, \bibinfo{author}{\bibfnamefont{X.}~\bibnamefont{Feng}}, \bibinfo{author}{\bibfnamefont{J.}~\bibnamefont{Shen}}, \bibinfo{author}{\bibfnamefont{Z.}~\bibnamefont{Zhang}}, \bibinfo{author}{\bibfnamefont{M.}~\bibnamefont{Guo}}, \bibinfo{author}{\bibfnamefont{K.}~\bibnamefont{Li}}, \bibinfo{author}{\bibfnamefont{Y.}~\bibnamefont{Ou}}, \bibinfo{author}{\bibfnamefont{P.}~\bibnamefont{Wei}}, \bibinfo{author}{\bibfnamefont{L.-L.} \bibnamefont{Wang}}, \bibnamefont{et~al.}, \bibinfo{journal}{Science} \textbf{\bibinfo{volume}{340}}, \bibinfo{pages}{167} (\bibinfo{year}{2013}{\natexlab{a}}), \eprint{https://www.science.org/doi/pdf/10.1126/science.1234414}, \urlprefix\url{https://www.science.org/doi/abs/10.1126/science.1234414}.

\bibitem[{\citenamefont{Kou et~al.}(2014)\citenamefont{Kou, Guo, Fan, Pan, Lang, Jiang, Shao, Nie, Murata, Tang et~al.}}]{kou_PhysRevLett.113.137201_2014}
\bibinfo{author}{\bibfnamefont{X.}~\bibnamefont{Kou}}, \bibinfo{author}{\bibfnamefont{S.-T.} \bibnamefont{Guo}}, \bibinfo{author}{\bibfnamefont{Y.}~\bibnamefont{Fan}}, \bibinfo{author}{\bibfnamefont{L.}~\bibnamefont{Pan}}, \bibinfo{author}{\bibfnamefont{M.}~\bibnamefont{Lang}}, \bibinfo{author}{\bibfnamefont{Y.}~\bibnamefont{Jiang}}, \bibinfo{author}{\bibfnamefont{Q.}~\bibnamefont{Shao}}, \bibinfo{author}{\bibfnamefont{T.}~\bibnamefont{Nie}}, \bibinfo{author}{\bibfnamefont{K.}~\bibnamefont{Murata}}, \bibinfo{author}{\bibfnamefont{J.}~\bibnamefont{Tang}}, \bibnamefont{et~al.}, \bibinfo{journal}{Phys. Rev. Lett.} \textbf{\bibinfo{volume}{113}}, \bibinfo{pages}{137201} (\bibinfo{year}{2014}), \urlprefix\url{https://link.aps.org/doi/10.1103/PhysRevLett.113.137201}.

\bibitem[{\citenamefont{Checkelsky et~al.}(2014)\citenamefont{Checkelsky, Yoshimi, Tsukazaki, Takahashi, Kozuka, Falson, Kawasaki, and Tokura}}]{checkelsky_natphy.10.731_2014}
\bibinfo{author}{\bibfnamefont{J.}~\bibnamefont{Checkelsky}}, \bibinfo{author}{\bibfnamefont{R.}~\bibnamefont{Yoshimi}}, \bibinfo{author}{\bibfnamefont{A.}~\bibnamefont{Tsukazaki}}, \bibinfo{author}{\bibfnamefont{K.}~\bibnamefont{Takahashi}}, \bibinfo{author}{\bibfnamefont{Y.}~\bibnamefont{Kozuka}}, \bibinfo{author}{\bibfnamefont{J.}~\bibnamefont{Falson}}, \bibinfo{author}{\bibfnamefont{M.}~\bibnamefont{Kawasaki}}, \bibnamefont{and} \bibinfo{author}{\bibfnamefont{Y.}~\bibnamefont{Tokura}}, \bibinfo{journal}{Nature Physics} \textbf{\bibinfo{volume}{10}}, \bibinfo{pages}{731} (\bibinfo{year}{2014}).

\bibitem[{\citenamefont{Bestwick et~al.}(2015)\citenamefont{Bestwick, Fox, Kou, Pan, Wang, and Goldhaber-Gordon}}]{bestwick_PhysRevLett.114.187201_2015}
\bibinfo{author}{\bibfnamefont{A.~J.} \bibnamefont{Bestwick}}, \bibinfo{author}{\bibfnamefont{E.~J.} \bibnamefont{Fox}}, \bibinfo{author}{\bibfnamefont{X.}~\bibnamefont{Kou}}, \bibinfo{author}{\bibfnamefont{L.}~\bibnamefont{Pan}}, \bibinfo{author}{\bibfnamefont{K.~L.} \bibnamefont{Wang}}, \bibnamefont{and} \bibinfo{author}{\bibfnamefont{D.}~\bibnamefont{Goldhaber-Gordon}}, \bibinfo{journal}{Phys. Rev. Lett.} \textbf{\bibinfo{volume}{114}}, \bibinfo{pages}{187201} (\bibinfo{year}{2015}), \urlprefix\url{https://link.aps.org/doi/10.1103/PhysRevLett.114.187201}.

\bibitem[{\citenamefont{Chang et~al.}(2015)\citenamefont{Chang, Zhao, Kim, Zhang, Assaf, Heiman, Zhang, Liu, Chan, and Moodera}}]{chang_natphy.14.473_2015}
\bibinfo{author}{\bibfnamefont{C.-Z.} \bibnamefont{Chang}}, \bibinfo{author}{\bibfnamefont{W.}~\bibnamefont{Zhao}}, \bibinfo{author}{\bibfnamefont{D.~Y.} \bibnamefont{Kim}}, \bibinfo{author}{\bibfnamefont{H.}~\bibnamefont{Zhang}}, \bibinfo{author}{\bibfnamefont{B.~A.} \bibnamefont{Assaf}}, \bibinfo{author}{\bibfnamefont{D.}~\bibnamefont{Heiman}}, \bibinfo{author}{\bibfnamefont{S.-C.} \bibnamefont{Zhang}}, \bibinfo{author}{\bibfnamefont{C.}~\bibnamefont{Liu}}, \bibinfo{author}{\bibfnamefont{M.~H.} \bibnamefont{Chan}}, \bibnamefont{and} \bibinfo{author}{\bibfnamefont{J.~S.} \bibnamefont{Moodera}}, \bibinfo{journal}{Nature materials} \textbf{\bibinfo{volume}{14}}, \bibinfo{pages}{473} (\bibinfo{year}{2015}).

\bibitem[{\citenamefont{Chen et~al.}(2018)\citenamefont{Chen, Xie, Liu, Lee, and Law}}]{chen_PhysRevB.97.104504_2018}
\bibinfo{author}{\bibfnamefont{C.-Z.} \bibnamefont{Chen}}, \bibinfo{author}{\bibfnamefont{Y.-M.} \bibnamefont{Xie}}, \bibinfo{author}{\bibfnamefont{J.}~\bibnamefont{Liu}}, \bibinfo{author}{\bibfnamefont{P.~A.} \bibnamefont{Lee}}, \bibnamefont{and} \bibinfo{author}{\bibfnamefont{K.~T.} \bibnamefont{Law}}, \bibinfo{journal}{Phys. Rev. B} \textbf{\bibinfo{volume}{97}}, \bibinfo{pages}{104504} (\bibinfo{year}{2018}), \urlprefix\url{https://link.aps.org/doi/10.1103/PhysRevB.97.104504}.

\bibitem[{\citenamefont{Feng et~al.}(2015)\citenamefont{Feng, Feng, Ou, Wang, Liu, Zhang, Zhao, Jiang, Zhang, He et~al.}}]{YFeng_PRL_115_2015}
\bibinfo{author}{\bibfnamefont{Y.}~\bibnamefont{Feng}}, \bibinfo{author}{\bibfnamefont{X.}~\bibnamefont{Feng}}, \bibinfo{author}{\bibfnamefont{Y.}~\bibnamefont{Ou}}, \bibinfo{author}{\bibfnamefont{J.}~\bibnamefont{Wang}}, \bibinfo{author}{\bibfnamefont{C.}~\bibnamefont{Liu}}, \bibinfo{author}{\bibfnamefont{L.}~\bibnamefont{Zhang}}, \bibinfo{author}{\bibfnamefont{D.}~\bibnamefont{Zhao}}, \bibinfo{author}{\bibfnamefont{G.}~\bibnamefont{Jiang}}, \bibinfo{author}{\bibfnamefont{S.-C.} \bibnamefont{Zhang}}, \bibinfo{author}{\bibfnamefont{K.}~\bibnamefont{He}}, \bibnamefont{et~al.}, \bibinfo{journal}{Phys. Rev. Lett.} \textbf{\bibinfo{volume}{115}}, \bibinfo{pages}{126801} (\bibinfo{year}{2015}), \urlprefix\url{https://link.aps.org/doi/10.1103/PhysRevLett.115.126801}.

\bibitem[{\citenamefont{Chang et~al.}(2023)\citenamefont{Chang, Liu, and MacDonald}}]{CChang_RevModPhys_95_2023}
\bibinfo{author}{\bibfnamefont{C.-Z.} \bibnamefont{Chang}}, \bibinfo{author}{\bibfnamefont{C.-X.} \bibnamefont{Liu}}, \bibnamefont{and} \bibinfo{author}{\bibfnamefont{A.~H.} \bibnamefont{MacDonald}}, \bibinfo{journal}{Rev. Mod. Phys.} \textbf{\bibinfo{volume}{95}}, \bibinfo{pages}{011002} (\bibinfo{year}{2023}), \urlprefix\url{https://link.aps.org/doi/10.1103/RevModPhys.95.011002}.

\bibitem[{\citenamefont{Ou et~al.}(2018)\citenamefont{Ou, Liu, Jiang, Feng, Zhao, Wu, Wang, Li, Song, Wang et~al.}}]{OYunbo_AdvMater_30_2017}
\bibinfo{author}{\bibfnamefont{Y.}~\bibnamefont{Ou}}, \bibinfo{author}{\bibfnamefont{C.}~\bibnamefont{Liu}}, \bibinfo{author}{\bibfnamefont{G.}~\bibnamefont{Jiang}}, \bibinfo{author}{\bibfnamefont{Y.}~\bibnamefont{Feng}}, \bibinfo{author}{\bibfnamefont{D.}~\bibnamefont{Zhao}}, \bibinfo{author}{\bibfnamefont{W.}~\bibnamefont{Wu}}, \bibinfo{author}{\bibfnamefont{X.-X.} \bibnamefont{Wang}}, \bibinfo{author}{\bibfnamefont{W.}~\bibnamefont{Li}}, \bibinfo{author}{\bibfnamefont{C.}~\bibnamefont{Song}}, \bibinfo{author}{\bibfnamefont{L.-L.} \bibnamefont{Wang}}, \bibnamefont{et~al.}, \bibinfo{journal}{Advanced Materials} \textbf{\bibinfo{volume}{30}}, \bibinfo{pages}{1703062} (\bibinfo{year}{2018}), \eprint{https://onlinelibrary.wiley.com/doi/pdf/10.1002/adma.201703062}, \urlprefix\url{https://onlinelibrary.wiley.com/doi/abs/10.1002/adma.201703062}.

\bibitem[{\citenamefont{Mogi et~al.}(2015)\citenamefont{Mogi, Yoshimi, Tsukazaki, Yasuda, Kozuka, Takahashi, Kawasaki, and Tokura}}]{MMogi_APL_107_2015}
\bibinfo{author}{\bibfnamefont{M.}~\bibnamefont{Mogi}}, \bibinfo{author}{\bibfnamefont{R.}~\bibnamefont{Yoshimi}}, \bibinfo{author}{\bibfnamefont{A.}~\bibnamefont{Tsukazaki}}, \bibinfo{author}{\bibfnamefont{K.}~\bibnamefont{Yasuda}}, \bibinfo{author}{\bibfnamefont{Y.}~\bibnamefont{Kozuka}}, \bibinfo{author}{\bibfnamefont{K.~S.} \bibnamefont{Takahashi}}, \bibinfo{author}{\bibfnamefont{M.}~\bibnamefont{Kawasaki}}, \bibnamefont{and} \bibinfo{author}{\bibfnamefont{Y.}~\bibnamefont{Tokura}}, \bibinfo{journal}{Applied Physics Letters} \textbf{\bibinfo{volume}{107}}, \bibinfo{pages}{182401} (\bibinfo{year}{2015}), \eprint{https://doi.org/10.1063/1.4935075}, \urlprefix\url{https://doi.org/10.1063/1.4935075}.

\bibitem[{\citenamefont{Zhou et~al.}(2005)\citenamefont{Zhou, Chien, and Uher}}]{ZZhou_APL_87_2005}
\bibinfo{author}{\bibfnamefont{Z.}~\bibnamefont{Zhou}}, \bibinfo{author}{\bibfnamefont{Y.-J.} \bibnamefont{Chien}}, \bibnamefont{and} \bibinfo{author}{\bibfnamefont{C.}~\bibnamefont{Uher}}, \bibinfo{journal}{Applied Physics Letters} \textbf{\bibinfo{volume}{87}}, \bibinfo{pages}{112503} (\bibinfo{year}{2005}), \eprint{https://doi.org/10.1063/1.2045561}, \urlprefix\url{https://doi.org/10.1063/1.2045561}.

\bibitem[{\citenamefont{Zhou et~al.}(2006)\citenamefont{Zhou, Chien, and Uher}}]{ZZhou_PRB_74_2006}
\bibinfo{author}{\bibfnamefont{Z.}~\bibnamefont{Zhou}}, \bibinfo{author}{\bibfnamefont{Y.-J.} \bibnamefont{Chien}}, \bibnamefont{and} \bibinfo{author}{\bibfnamefont{C.}~\bibnamefont{Uher}}, \bibinfo{journal}{Phys. Rev. B} \textbf{\bibinfo{volume}{74}}, \bibinfo{pages}{224418} (\bibinfo{year}{2006}), \urlprefix\url{https://link.aps.org/doi/10.1103/PhysRevB.74.224418}.

\bibitem[{\citenamefont{Werpachowska and Wilamowski}(2011)}]{AWerpachowska_arxiv.1111.2011}
\bibinfo{author}{\bibfnamefont{A.}~\bibnamefont{Werpachowska}} \bibnamefont{and} \bibinfo{author}{\bibfnamefont{Z.}~\bibnamefont{Wilamowski}}, \emph{\bibinfo{title}{The rkky coupling in diluted magnetic semiconductors}} (\bibinfo{year}{2011}), \eprint{1111.1030}.

\bibitem[{\citenamefont{Dietl et~al.}(1997)\citenamefont{Dietl, Haury, and Merle~d'Aubign\'e}}]{TDietl_PRB.55_1997}
\bibinfo{author}{\bibfnamefont{T.}~\bibnamefont{Dietl}}, \bibinfo{author}{\bibfnamefont{A.}~\bibnamefont{Haury}}, \bibnamefont{and} \bibinfo{author}{\bibfnamefont{Y.}~\bibnamefont{Merle~d'Aubign\'e}}, \bibinfo{journal}{Phys. Rev. B} \textbf{\bibinfo{volume}{55}}, \bibinfo{pages}{R3347} (\bibinfo{year}{1997}), \urlprefix\url{https://link.aps.org/doi/10.1103/PhysRevB.55.R3347}.

\bibitem[{\citenamefont{Yu et~al.}(2010{\natexlab{b}})\citenamefont{Yu, Zhang, Zhang, Zhang, Dai, and Fang}}]{YRui_science.329_2010}
\bibinfo{author}{\bibfnamefont{R.}~\bibnamefont{Yu}}, \bibinfo{author}{\bibfnamefont{W.}~\bibnamefont{Zhang}}, \bibinfo{author}{\bibfnamefont{H.-J.} \bibnamefont{Zhang}}, \bibinfo{author}{\bibfnamefont{S.-C.} \bibnamefont{Zhang}}, \bibinfo{author}{\bibfnamefont{X.}~\bibnamefont{Dai}}, \bibnamefont{and} \bibinfo{author}{\bibfnamefont{Z.}~\bibnamefont{Fang}}, \bibinfo{journal}{Science} \textbf{\bibinfo{volume}{329}}, \bibinfo{pages}{61} (\bibinfo{year}{2010}{\natexlab{b}}), \eprint{https://www.science.org/doi/pdf/10.1126/science.1187485}, \urlprefix\url{https://www.science.org/doi/abs/10.1126/science.1187485}.

\bibitem[{\citenamefont{Van~Vlek}(2015)}]{VVlek_book_2015}
\bibinfo{author}{\bibfnamefont{J.}~\bibnamefont{Van~Vlek}}, \emph{\bibinfo{title}{The Theory of Electric and Magnetic Susceptibilities - Scholar's Choice Edition}} (\bibinfo{publisher}{Bibliolife DBA of Bibilio Bazaar II LLC}, \bibinfo{year}{2015}), ISBN \bibinfo{isbn}{9781298031464}, \urlprefix\url{https://books.google.se/books?id=C5o2rgEACAAJ}.

\bibitem[{\citenamefont{Chang et~al.}(2013{\natexlab{b}})\citenamefont{Chang, Zhang, Liu, Zhang, Feng, Li, Wang, Chen, Dai, Fang et~al.}}]{CCui_AdvMat.25_2013}
\bibinfo{author}{\bibfnamefont{C.-Z.} \bibnamefont{Chang}}, \bibinfo{author}{\bibfnamefont{J.}~\bibnamefont{Zhang}}, \bibinfo{author}{\bibfnamefont{M.}~\bibnamefont{Liu}}, \bibinfo{author}{\bibfnamefont{Z.}~\bibnamefont{Zhang}}, \bibinfo{author}{\bibfnamefont{X.}~\bibnamefont{Feng}}, \bibinfo{author}{\bibfnamefont{K.}~\bibnamefont{Li}}, \bibinfo{author}{\bibfnamefont{L.-L.} \bibnamefont{Wang}}, \bibinfo{author}{\bibfnamefont{X.}~\bibnamefont{Chen}}, \bibinfo{author}{\bibfnamefont{X.}~\bibnamefont{Dai}}, \bibinfo{author}{\bibfnamefont{Z.}~\bibnamefont{Fang}}, \bibnamefont{et~al.}, \bibinfo{journal}{Advanced Materials} \textbf{\bibinfo{volume}{25}}, \bibinfo{pages}{1065} (\bibinfo{year}{2013}{\natexlab{b}}), \eprint{https://onlinelibrary.wiley.com/doi/pdf/10.1002/adma.201203493}, \urlprefix\url{https://onlinelibrary.wiley.com/doi/abs/10.1002/adma.201203493}.

\bibitem[{\citenamefont{Li et~al.}(2015)\citenamefont{Li, Chang, Wu, Tao, Zhao, Chan, Moodera, Li, and Zhu}}]{LMingda_PRL.114_2015}
\bibinfo{author}{\bibfnamefont{M.}~\bibnamefont{Li}}, \bibinfo{author}{\bibfnamefont{C.-Z.} \bibnamefont{Chang}}, \bibinfo{author}{\bibfnamefont{L.}~\bibnamefont{Wu}}, \bibinfo{author}{\bibfnamefont{J.}~\bibnamefont{Tao}}, \bibinfo{author}{\bibfnamefont{W.}~\bibnamefont{Zhao}}, \bibinfo{author}{\bibfnamefont{M.~H.~W.} \bibnamefont{Chan}}, \bibinfo{author}{\bibfnamefont{J.~S.} \bibnamefont{Moodera}}, \bibinfo{author}{\bibfnamefont{J.}~\bibnamefont{Li}}, \bibnamefont{and} \bibinfo{author}{\bibfnamefont{Y.}~\bibnamefont{Zhu}}, \bibinfo{journal}{Phys. Rev. Lett.} \textbf{\bibinfo{volume}{114}}, \bibinfo{pages}{146802} (\bibinfo{year}{2015}), \urlprefix\url{https://link.aps.org/doi/10.1103/PhysRevLett.114.146802}.

\bibitem[{\citenamefont{Vergniory et~al.}(2014)\citenamefont{Vergniory, Otrokov, Thonig, Hoffmann, Maznichenko, Geilhufe, Zubizarreta, Ostanin, Marmodoro, Henk et~al.}}]{verginory_PhysRevB}
\bibinfo{author}{\bibfnamefont{M.~G.} \bibnamefont{Vergniory}}, \bibinfo{author}{\bibfnamefont{M.~M.} \bibnamefont{Otrokov}}, \bibinfo{author}{\bibfnamefont{D.}~\bibnamefont{Thonig}}, \bibinfo{author}{\bibfnamefont{M.}~\bibnamefont{Hoffmann}}, \bibinfo{author}{\bibfnamefont{I.~V.} \bibnamefont{Maznichenko}}, \bibinfo{author}{\bibfnamefont{M.}~\bibnamefont{Geilhufe}}, \bibinfo{author}{\bibfnamefont{X.}~\bibnamefont{Zubizarreta}}, \bibinfo{author}{\bibfnamefont{S.}~\bibnamefont{Ostanin}}, \bibinfo{author}{\bibfnamefont{A.}~\bibnamefont{Marmodoro}}, \bibinfo{author}{\bibfnamefont{J.}~\bibnamefont{Henk}}, \bibnamefont{et~al.}, \bibinfo{journal}{Phys. Rev. B} \textbf{\bibinfo{volume}{89}}, \bibinfo{pages}{165202} (\bibinfo{year}{2014}), \urlprefix\url{https://link.aps.org/doi/10.1103/PhysRevB.89.165202}.

\bibitem[{\citenamefont{Rüßmann et~al.}(2018)\citenamefont{Rüßmann, Mahatha, Sessi, Valbuena, Bathon, Fauth, Godey, Mugarza, Kokh, Tereshchenko et~al.}}]{PRubmann_JPM.1_2018}
\bibinfo{author}{\bibfnamefont{P.}~\bibnamefont{Rüßmann}}, \bibinfo{author}{\bibfnamefont{S.~K.} \bibnamefont{Mahatha}}, \bibinfo{author}{\bibfnamefont{P.}~\bibnamefont{Sessi}}, \bibinfo{author}{\bibfnamefont{M.~A.} \bibnamefont{Valbuena}}, \bibinfo{author}{\bibfnamefont{T.}~\bibnamefont{Bathon}}, \bibinfo{author}{\bibfnamefont{K.}~\bibnamefont{Fauth}}, \bibinfo{author}{\bibfnamefont{S.}~\bibnamefont{Godey}}, \bibinfo{author}{\bibfnamefont{A.}~\bibnamefont{Mugarza}}, \bibinfo{author}{\bibfnamefont{K.~A.} \bibnamefont{Kokh}}, \bibinfo{author}{\bibfnamefont{O.~E.} \bibnamefont{Tereshchenko}}, \bibnamefont{et~al.}, \bibinfo{journal}{Journal of Physics: Materials} \textbf{\bibinfo{volume}{1}}, \bibinfo{pages}{015002} (\bibinfo{year}{2018}), \urlprefix\url{https://dx.doi.org/10.1088/2515-7639/aad02a}.

\bibitem[{\citenamefont{Zhang et~al.}(2014)\citenamefont{Zhang, Feng, Guo, Li, Zhang, Ou, Feng, Wang, Chen, He et~al.}}]{ZZhang_Natcom.5_2014}
\bibinfo{author}{\bibfnamefont{Z.}~\bibnamefont{Zhang}}, \bibinfo{author}{\bibfnamefont{X.}~\bibnamefont{Feng}}, \bibinfo{author}{\bibfnamefont{M.}~\bibnamefont{Guo}}, \bibinfo{author}{\bibfnamefont{K.}~\bibnamefont{Li}}, \bibinfo{author}{\bibfnamefont{J.}~\bibnamefont{Zhang}}, \bibinfo{author}{\bibfnamefont{Y.}~\bibnamefont{Ou}}, \bibinfo{author}{\bibfnamefont{Y.}~\bibnamefont{Feng}}, \bibinfo{author}{\bibfnamefont{L.}~\bibnamefont{Wang}}, \bibinfo{author}{\bibfnamefont{X.}~\bibnamefont{Chen}}, \bibinfo{author}{\bibfnamefont{K.}~\bibnamefont{He}}, \bibnamefont{et~al.}, \bibinfo{journal}{Nature communications} \textbf{\bibinfo{volume}{5}}, \bibinfo{pages}{4915} (\bibinfo{year}{2014}).

\bibitem[{\citenamefont{Wang et~al.}(2023)\citenamefont{Wang, Zhao, Yan, Zhuo, Yi, Yuan, Zhou, Zhao, Chan, and Chang}}]{FWang_NanoLett.23_2023}
\bibinfo{author}{\bibfnamefont{F.}~\bibnamefont{Wang}}, \bibinfo{author}{\bibfnamefont{Y.-F.} \bibnamefont{Zhao}}, \bibinfo{author}{\bibfnamefont{Z.-J.} \bibnamefont{Yan}}, \bibinfo{author}{\bibfnamefont{D.}~\bibnamefont{Zhuo}}, \bibinfo{author}{\bibfnamefont{H.}~\bibnamefont{Yi}}, \bibinfo{author}{\bibfnamefont{W.}~\bibnamefont{Yuan}}, \bibinfo{author}{\bibfnamefont{L.}~\bibnamefont{Zhou}}, \bibinfo{author}{\bibfnamefont{W.}~\bibnamefont{Zhao}}, \bibinfo{author}{\bibfnamefont{M.~H.~W.} \bibnamefont{Chan}}, \bibnamefont{and} \bibinfo{author}{\bibfnamefont{C.-Z.} \bibnamefont{Chang}}, \bibinfo{journal}{Nano Letters} \textbf{\bibinfo{volume}{23}}, \bibinfo{pages}{2483} (\bibinfo{year}{2023}), \bibinfo{note}{pMID: 36930727}, \eprint{https://doi.org/10.1021/acs.nanolett.2c03827}, \urlprefix\url{https://doi.org/10.1021/acs.nanolett.2c03827}.

\bibitem[{\citenamefont{Kim et~al.}(2017)\citenamefont{Kim, Jhi, MacDonald, and Wu}}]{Jkim_PRB.96_2017}
\bibinfo{author}{\bibfnamefont{J.}~\bibnamefont{Kim}}, \bibinfo{author}{\bibfnamefont{S.-H.} \bibnamefont{Jhi}}, \bibinfo{author}{\bibfnamefont{A.~H.} \bibnamefont{MacDonald}}, \bibnamefont{and} \bibinfo{author}{\bibfnamefont{R.}~\bibnamefont{Wu}}, \bibinfo{journal}{Phys. Rev. B} \textbf{\bibinfo{volume}{96}}, \bibinfo{pages}{140410} (\bibinfo{year}{2017}), \urlprefix\url{https://link.aps.org/doi/10.1103/PhysRevB.96.140410}.

\bibitem[{\citenamefont{Kim et~al.}(2018)\citenamefont{Kim, Wang, and Wu}}]{Jkim_PRB.97_2018}
\bibinfo{author}{\bibfnamefont{J.}~\bibnamefont{Kim}}, \bibinfo{author}{\bibfnamefont{H.}~\bibnamefont{Wang}}, \bibnamefont{and} \bibinfo{author}{\bibfnamefont{R.}~\bibnamefont{Wu}}, \bibinfo{journal}{Phys. Rev. B} \textbf{\bibinfo{volume}{97}}, \bibinfo{pages}{125118} (\bibinfo{year}{2018}), \urlprefix\url{https://link.aps.org/doi/10.1103/PhysRevB.97.125118}.

\bibitem[{\citenamefont{Peixoto et~al.}(2020{\natexlab{a}})\citenamefont{Peixoto, Bentmann, R{\"u}{\ss}mann, Tcakaev, Winnerlein, Schreyeck, Schatz, Vidal, Stier, Zabolotnyy et~al.}}]{TPeixoto_npjQM.5_2020}
\bibinfo{author}{\bibfnamefont{T.~R.} \bibnamefont{Peixoto}}, \bibinfo{author}{\bibfnamefont{H.}~\bibnamefont{Bentmann}}, \bibinfo{author}{\bibfnamefont{P.}~\bibnamefont{R{\"u}{\ss}mann}}, \bibinfo{author}{\bibfnamefont{A.-V.} \bibnamefont{Tcakaev}}, \bibinfo{author}{\bibfnamefont{M.}~\bibnamefont{Winnerlein}}, \bibinfo{author}{\bibfnamefont{S.}~\bibnamefont{Schreyeck}}, \bibinfo{author}{\bibfnamefont{S.}~\bibnamefont{Schatz}}, \bibinfo{author}{\bibfnamefont{R.~C.} \bibnamefont{Vidal}}, \bibinfo{author}{\bibfnamefont{F.}~\bibnamefont{Stier}}, \bibinfo{author}{\bibfnamefont{V.}~\bibnamefont{Zabolotnyy}}, \bibnamefont{et~al.}, \bibinfo{journal}{npj quantum materials} \textbf{\bibinfo{volume}{5}}, \bibinfo{pages}{87} (\bibinfo{year}{2020}{\natexlab{a}}).

\bibitem[{\citenamefont{Kim and Jhi}(2015)}]{Jkim_PRB.92_2015}
\bibinfo{author}{\bibfnamefont{J.}~\bibnamefont{Kim}} \bibnamefont{and} \bibinfo{author}{\bibfnamefont{S.-H.} \bibnamefont{Jhi}}, \bibinfo{journal}{Phys. Rev. B} \textbf{\bibinfo{volume}{92}}, \bibinfo{pages}{104405} (\bibinfo{year}{2015}), \urlprefix\url{https://link.aps.org/doi/10.1103/PhysRevB.92.104405}.

\bibitem[{\citenamefont{Bl\"ochl}(1994)}]{blochl_PhysRevB.50.17953_1994}
\bibinfo{author}{\bibfnamefont{P.~E.} \bibnamefont{Bl\"ochl}}, \bibinfo{journal}{Phys. Rev. B} \textbf{\bibinfo{volume}{50}}, \bibinfo{pages}{17953} (\bibinfo{year}{1994}), \urlprefix\url{https://link.aps.org/doi/10.1103/PhysRevB.50.17953}.

\bibitem[{\citenamefont{Kresse and Joubert}(1999)}]{kresse_PhysRevB.59.1758_1999}
\bibinfo{author}{\bibfnamefont{G.}~\bibnamefont{Kresse}} \bibnamefont{and} \bibinfo{author}{\bibfnamefont{D.}~\bibnamefont{Joubert}}, \bibinfo{journal}{Phys. Rev. B} \textbf{\bibinfo{volume}{59}}, \bibinfo{pages}{1758} (\bibinfo{year}{1999}), \urlprefix\url{https://link.aps.org/doi/10.1103/PhysRevB.59.1758}.

\bibitem[{\citenamefont{Kresse and Hafner}(1993)}]{kresse_PhysRevB.47.558_1993}
\bibinfo{author}{\bibfnamefont{G.}~\bibnamefont{Kresse}} \bibnamefont{and} \bibinfo{author}{\bibfnamefont{J.}~\bibnamefont{Hafner}}, \bibinfo{journal}{Phys. Rev. B} \textbf{\bibinfo{volume}{47}}, \bibinfo{pages}{558} (\bibinfo{year}{1993}), \urlprefix\url{https://link.aps.org/doi/10.1103/PhysRevB.47.558}.

\bibitem[{\citenamefont{Kresse and Hafner}(1994)}]{kresse_PhysRevB.49.14251_1994}
\bibinfo{author}{\bibfnamefont{G.}~\bibnamefont{Kresse}} \bibnamefont{and} \bibinfo{author}{\bibfnamefont{J.}~\bibnamefont{Hafner}}, \bibinfo{journal}{Phys. Rev. B} \textbf{\bibinfo{volume}{49}}, \bibinfo{pages}{14251} (\bibinfo{year}{1994}), \urlprefix\url{https://link.aps.org/doi/10.1103/PhysRevB.49.14251}.

\bibitem[{\citenamefont{Kresse and Furthm\"uller}(1996)}]{kresse_PhysRevB.54.11169_1996}
\bibinfo{author}{\bibfnamefont{G.}~\bibnamefont{Kresse}} \bibnamefont{and} \bibinfo{author}{\bibfnamefont{J.}~\bibnamefont{Furthm\"uller}}, \bibinfo{journal}{Phys. Rev. B} \textbf{\bibinfo{volume}{54}}, \bibinfo{pages}{11169} (\bibinfo{year}{1996}), \urlprefix\url{https://link.aps.org/doi/10.1103/PhysRevB.54.11169}.

\bibitem[{\citenamefont{Kresse and Furthmüller}(1996)}]{kresse_cms.6.15_1996}
\bibinfo{author}{\bibfnamefont{G.}~\bibnamefont{Kresse}} \bibnamefont{and} \bibinfo{author}{\bibfnamefont{J.}~\bibnamefont{Furthmüller}}, \bibinfo{journal}{Computational Materials Science} \textbf{\bibinfo{volume}{6}}, \bibinfo{pages}{15} (\bibinfo{year}{1996}), ISSN \bibinfo{issn}{0927-0256}, \urlprefix\url{https://www.sciencedirect.com/science/article/pii/0927025696000080}.

\bibitem[{\citenamefont{Dirac}(1930)}]{dirac_chembridge.26.376_1930}
\bibinfo{author}{\bibfnamefont{P.~A.~M.} \bibnamefont{Dirac}}, \bibinfo{journal}{Mathematical Proceedings of the Cambridge Philosophical Society} \textbf{\bibinfo{volume}{26}}, \bibinfo{pages}{376–385} (\bibinfo{year}{1930}).

\bibitem[{\citenamefont{Ceperley and Alder}(1980)}]{ceperley_PhysRevLett.45.566_1980}
\bibinfo{author}{\bibfnamefont{D.~M.} \bibnamefont{Ceperley}} \bibnamefont{and} \bibinfo{author}{\bibfnamefont{B.~J.} \bibnamefont{Alder}}, \bibinfo{journal}{Phys. Rev. Lett.} \textbf{\bibinfo{volume}{45}}, \bibinfo{pages}{566} (\bibinfo{year}{1980}), \urlprefix\url{https://link.aps.org/doi/10.1103/PhysRevLett.45.566}.

\bibitem[{\citenamefont{Perdew and Zunger}(1981)}]{perdew_PhysRevB.23.5048_1981}
\bibinfo{author}{\bibfnamefont{J.~P.} \bibnamefont{Perdew}} \bibnamefont{and} \bibinfo{author}{\bibfnamefont{A.}~\bibnamefont{Zunger}}, \bibinfo{journal}{Phys. Rev. B} \textbf{\bibinfo{volume}{23}}, \bibinfo{pages}{5048} (\bibinfo{year}{1981}), \urlprefix\url{https://link.aps.org/doi/10.1103/PhysRevB.23.5048}.

\bibitem[{\citenamefont{Perdew et~al.}(1996)\citenamefont{Perdew, Burke, and Ernzerhof}}]{perdew_PhysRevLett.77.3865_1996}
\bibinfo{author}{\bibfnamefont{J.~P.} \bibnamefont{Perdew}}, \bibinfo{author}{\bibfnamefont{K.}~\bibnamefont{Burke}}, \bibnamefont{and} \bibinfo{author}{\bibfnamefont{M.}~\bibnamefont{Ernzerhof}}, \bibinfo{journal}{Phys. Rev. Lett.} \textbf{\bibinfo{volume}{77}}, \bibinfo{pages}{3865} (\bibinfo{year}{1996}), \urlprefix\url{https://link.aps.org/doi/10.1103/PhysRevLett.77.3865}.

\bibitem[{\citenamefont{Perdew et~al.}(1997)\citenamefont{Perdew, Burke, and Ernzerhof}}]{perdew_PhysRevLett.78.1396_1997}
\bibinfo{author}{\bibfnamefont{J.~P.} \bibnamefont{Perdew}}, \bibinfo{author}{\bibfnamefont{K.}~\bibnamefont{Burke}}, \bibnamefont{and} \bibinfo{author}{\bibfnamefont{M.}~\bibnamefont{Ernzerhof}}, \bibinfo{journal}{Phys. Rev. Lett.} \textbf{\bibinfo{volume}{78}}, \bibinfo{pages}{1396} (\bibinfo{year}{1997}), \urlprefix\url{https://link.aps.org/doi/10.1103/PhysRevLett.78.1396}.

\bibitem[{\citenamefont{Grimme}(2006)}]{grimme_jcc.27.15_2006}
\bibinfo{author}{\bibfnamefont{S.}~\bibnamefont{Grimme}}, \bibinfo{journal}{Journal of Computational Chemistry} \textbf{\bibinfo{volume}{27}}, \bibinfo{pages}{1787} (\bibinfo{year}{2006}), \eprint{https://onlinelibrary.wiley.com/doi/pdf/10.1002/jcc.20495}, \urlprefix\url{https://onlinelibrary.wiley.com/doi/abs/10.1002/jcc.20495}.

\bibitem[{\citenamefont{Steiner et~al.}(2016)\citenamefont{Steiner, Khmelevskyi, Marsmann, and Kresse}}]{steiner_PhysRevB.93.224425_2016}
\bibinfo{author}{\bibfnamefont{S.}~\bibnamefont{Steiner}}, \bibinfo{author}{\bibfnamefont{S.}~\bibnamefont{Khmelevskyi}}, \bibinfo{author}{\bibfnamefont{M.}~\bibnamefont{Marsmann}}, \bibnamefont{and} \bibinfo{author}{\bibfnamefont{G.}~\bibnamefont{Kresse}}, \bibinfo{journal}{Phys. Rev. B} \textbf{\bibinfo{volume}{93}}, \bibinfo{pages}{224425} (\bibinfo{year}{2016}), \urlprefix\url{https://link.aps.org/doi/10.1103/PhysRevB.93.224425}.

\bibitem[{\citenamefont{Anisimov et~al.}(1997{\natexlab{a}})\citenamefont{Anisimov, Platow, Poulopoulos, Wisny, Farle, Baberschke, Isberg, Hjörvarsson, and Wäppling}}]{anisimov_jpcm.9.48_1997}
\bibinfo{author}{\bibfnamefont{A.~N.} \bibnamefont{Anisimov}}, \bibinfo{author}{\bibfnamefont{W.}~\bibnamefont{Platow}}, \bibinfo{author}{\bibfnamefont{P.}~\bibnamefont{Poulopoulos}}, \bibinfo{author}{\bibfnamefont{W.}~\bibnamefont{Wisny}}, \bibinfo{author}{\bibfnamefont{M.}~\bibnamefont{Farle}}, \bibinfo{author}{\bibfnamefont{K.}~\bibnamefont{Baberschke}}, \bibinfo{author}{\bibfnamefont{P.}~\bibnamefont{Isberg}}, \bibinfo{author}{\bibfnamefont{B.}~\bibnamefont{Hjörvarsson}}, \bibnamefont{and} \bibinfo{author}{\bibfnamefont{R.}~\bibnamefont{Wäppling}}, \bibinfo{journal}{Journal of Physics: Condensed Matter} \textbf{\bibinfo{volume}{9}}, \bibinfo{pages}{10581} (\bibinfo{year}{1997}{\natexlab{a}}), \urlprefix\url{https://doi.org/10.1088/0953-8984/9/48/004}.

\bibitem[{\citenamefont{Kotliar et~al.}(2006{\natexlab{a}})\citenamefont{Kotliar, Savrasov, Haule, Oudovenko, Parcollet, and Marianetti}}]{kotliar_RevModPhys.78.865_2006}
\bibinfo{author}{\bibfnamefont{G.}~\bibnamefont{Kotliar}}, \bibinfo{author}{\bibfnamefont{S.~Y.} \bibnamefont{Savrasov}}, \bibinfo{author}{\bibfnamefont{K.}~\bibnamefont{Haule}}, \bibinfo{author}{\bibfnamefont{V.~S.} \bibnamefont{Oudovenko}}, \bibinfo{author}{\bibfnamefont{O.}~\bibnamefont{Parcollet}}, \bibnamefont{and} \bibinfo{author}{\bibfnamefont{C.~A.} \bibnamefont{Marianetti}}, \bibinfo{journal}{Rev. Mod. Phys.} \textbf{\bibinfo{volume}{78}}, \bibinfo{pages}{865} (\bibinfo{year}{2006}{\natexlab{a}}), \urlprefix\url{https://link.aps.org/doi/10.1103/RevModPhys.78.865}.

\bibitem[{\citenamefont{Liechtenstein et~al.}(1995)\citenamefont{Liechtenstein, Anisimov, and Zaanen}}]{liechtenstein_PhysRevB.52.R5467_1995}
\bibinfo{author}{\bibfnamefont{A.~I.} \bibnamefont{Liechtenstein}}, \bibinfo{author}{\bibfnamefont{V.~I.} \bibnamefont{Anisimov}}, \bibnamefont{and} \bibinfo{author}{\bibfnamefont{J.}~\bibnamefont{Zaanen}}, \bibinfo{journal}{Phys. Rev. B} \textbf{\bibinfo{volume}{52}}, \bibinfo{pages}{R5467} (\bibinfo{year}{1995}), \urlprefix\url{https://link.aps.org/doi/10.1103/PhysRevB.52.R5467}.

\bibitem[{\citenamefont{Islam et~al.}(2018)\citenamefont{Islam, Canali, Pertsova, Balatsky, Mahatha, Carbone, Barla, Kokh, Tereshchenko, Jim\'enez et~al.}}]{MFIslam_PRB.97_2018}
\bibinfo{author}{\bibfnamefont{M.~F.} \bibnamefont{Islam}}, \bibinfo{author}{\bibfnamefont{C.~M.} \bibnamefont{Canali}}, \bibinfo{author}{\bibfnamefont{A.}~\bibnamefont{Pertsova}}, \bibinfo{author}{\bibfnamefont{A.}~\bibnamefont{Balatsky}}, \bibinfo{author}{\bibfnamefont{S.~K.} \bibnamefont{Mahatha}}, \bibinfo{author}{\bibfnamefont{C.}~\bibnamefont{Carbone}}, \bibinfo{author}{\bibfnamefont{A.}~\bibnamefont{Barla}}, \bibinfo{author}{\bibfnamefont{K.~A.} \bibnamefont{Kokh}}, \bibinfo{author}{\bibfnamefont{O.~E.} \bibnamefont{Tereshchenko}}, \bibinfo{author}{\bibfnamefont{E.}~\bibnamefont{Jim\'enez}}, \bibnamefont{et~al.}, \bibinfo{journal}{Phys. Rev. B} \textbf{\bibinfo{volume}{97}}, \bibinfo{pages}{155429} (\bibinfo{year}{2018}), \urlprefix\url{https://link.aps.org/doi/10.1103/PhysRevB.97.155429}.

\bibitem[{\citenamefont{Yang et~al.}(2020)\citenamefont{Yang, Fan, Wang, Khomskii, and Wu}}]{Kyang_PRB.101_2020}
\bibinfo{author}{\bibfnamefont{K.}~\bibnamefont{Yang}}, \bibinfo{author}{\bibfnamefont{F.}~\bibnamefont{Fan}}, \bibinfo{author}{\bibfnamefont{H.}~\bibnamefont{Wang}}, \bibinfo{author}{\bibfnamefont{D.~I.} \bibnamefont{Khomskii}}, \bibnamefont{and} \bibinfo{author}{\bibfnamefont{H.}~\bibnamefont{Wu}}, \bibinfo{journal}{Phys. Rev. B} \textbf{\bibinfo{volume}{101}}, \bibinfo{pages}{100402} (\bibinfo{year}{2020}), \urlprefix\url{https://link.aps.org/doi/10.1103/PhysRevB.101.100402}.

\bibitem[{\citenamefont{Wills et~al.}(2010)\citenamefont{Wills, Alouani, Andersson, Delin, Eriksson, and Grechnyev}}]{wills2010full}
\bibinfo{author}{\bibfnamefont{J.~M.} \bibnamefont{Wills}}, \bibinfo{author}{\bibfnamefont{M.}~\bibnamefont{Alouani}}, \bibinfo{author}{\bibfnamefont{P.}~\bibnamefont{Andersson}}, \bibinfo{author}{\bibfnamefont{A.}~\bibnamefont{Delin}}, \bibinfo{author}{\bibfnamefont{O.}~\bibnamefont{Eriksson}}, \bibnamefont{and} \bibinfo{author}{\bibfnamefont{O.}~\bibnamefont{Grechnyev}}, \emph{\bibinfo{title}{Full-Potential Electronic Structure Method: energy and force calculations with density functional and dynamical mean field theory}}, vol. \bibinfo{volume}{167} (\bibinfo{publisher}{Springer Science \& Business Media}, \bibinfo{year}{2010}).

\bibitem[{RSP()}]{RSPt}
\bibinfo{howpublished}{\url{https://www.physics.uu.se/research/materials-theory/ongoing-research/code-development/rspt-main/}}, \bibinfo{note}{source RSPt}.

\bibitem[{\citenamefont{Grechnev et~al.}(2007)\citenamefont{Grechnev, Di~Marco, Katsnelson, Lichtenstein, Wills, and Eriksson}}]{PhysRevB_76}
\bibinfo{author}{\bibfnamefont{A.}~\bibnamefont{Grechnev}}, \bibinfo{author}{\bibfnamefont{I.}~\bibnamefont{Di~Marco}}, \bibinfo{author}{\bibfnamefont{M.~I.} \bibnamefont{Katsnelson}}, \bibinfo{author}{\bibfnamefont{A.~I.} \bibnamefont{Lichtenstein}}, \bibinfo{author}{\bibfnamefont{J.}~\bibnamefont{Wills}}, \bibnamefont{and} \bibinfo{author}{\bibfnamefont{O.}~\bibnamefont{Eriksson}}, \bibinfo{journal}{Phys. Rev. B} \textbf{\bibinfo{volume}{76}}, \bibinfo{pages}{035107} (\bibinfo{year}{2007}), \urlprefix\url{https://link.aps.org/doi/10.1103/PhysRevB.76.035107}.

\bibitem[{\citenamefont{Grånäs et~al.}(2012)\citenamefont{Grånäs, {Di Marco}, Thunström, Nordström, Eriksson, Björkman, and Wills}}]{GRANAS2012295}
\bibinfo{author}{\bibfnamefont{O.}~\bibnamefont{Grånäs}}, \bibinfo{author}{\bibfnamefont{I.}~\bibnamefont{{Di Marco}}}, \bibinfo{author}{\bibfnamefont{P.}~\bibnamefont{Thunström}}, \bibinfo{author}{\bibfnamefont{L.}~\bibnamefont{Nordström}}, \bibinfo{author}{\bibfnamefont{O.}~\bibnamefont{Eriksson}}, \bibinfo{author}{\bibfnamefont{T.}~\bibnamefont{Björkman}}, \bibnamefont{and} \bibinfo{author}{\bibfnamefont{J.}~\bibnamefont{Wills}}, \bibinfo{journal}{Computational Materials Science} \textbf{\bibinfo{volume}{55}}, \bibinfo{pages}{295} (\bibinfo{year}{2012}), ISSN \bibinfo{issn}{0927-0256}, \urlprefix\url{https://www.sciencedirect.com/science/article/pii/S092702561100646X}.

\bibitem[{\citenamefont{Kotliar et~al.}(2006{\natexlab{b}})\citenamefont{Kotliar, Savrasov, Haule, Oudovenko, Parcollet, and Marianetti}}]{RevModPhys_78}
\bibinfo{author}{\bibfnamefont{G.}~\bibnamefont{Kotliar}}, \bibinfo{author}{\bibfnamefont{S.~Y.} \bibnamefont{Savrasov}}, \bibinfo{author}{\bibfnamefont{K.}~\bibnamefont{Haule}}, \bibinfo{author}{\bibfnamefont{V.~S.} \bibnamefont{Oudovenko}}, \bibinfo{author}{\bibfnamefont{O.}~\bibnamefont{Parcollet}}, \bibnamefont{and} \bibinfo{author}{\bibfnamefont{C.~A.} \bibnamefont{Marianetti}}, \bibinfo{journal}{Rev. Mod. Phys.} \textbf{\bibinfo{volume}{78}}, \bibinfo{pages}{865} (\bibinfo{year}{2006}{\natexlab{b}}), \urlprefix\url{https://link.aps.org/doi/10.1103/RevModPhys.78.865}.

\bibitem[{\citenamefont{Anisimov et~al.}(1997{\natexlab{b}})\citenamefont{Anisimov, Aryasetiawan, and Lichtenstein}}]{Vladimir_1997}
\bibinfo{author}{\bibfnamefont{V.~I.} \bibnamefont{Anisimov}}, \bibinfo{author}{\bibfnamefont{F.}~\bibnamefont{Aryasetiawan}}, \bibnamefont{and} \bibinfo{author}{\bibfnamefont{A.~I.} \bibnamefont{Lichtenstein}}, \bibinfo{journal}{Journal of Physics: Condensed Matter} \textbf{\bibinfo{volume}{9}}, \bibinfo{pages}{767} (\bibinfo{year}{1997}{\natexlab{b}}), \urlprefix\url{https://dx.doi.org/10.1088/0953-8984/9/4/002}.

\bibitem[{\citenamefont{Liechtenstein et~al.}(1987)\citenamefont{Liechtenstein, Katsnelson, Antropov, and Gubanov}}]{LIECHTENSTEIN198765}
\bibinfo{author}{\bibfnamefont{A.}~\bibnamefont{Liechtenstein}}, \bibinfo{author}{\bibfnamefont{M.}~\bibnamefont{Katsnelson}}, \bibinfo{author}{\bibfnamefont{V.}~\bibnamefont{Antropov}}, \bibnamefont{and} \bibinfo{author}{\bibfnamefont{V.}~\bibnamefont{Gubanov}}, \bibinfo{journal}{Journal of Magnetism and Magnetic Materials} \textbf{\bibinfo{volume}{67}}, \bibinfo{pages}{65} (\bibinfo{year}{1987}), ISSN \bibinfo{issn}{0304-8853}, \urlprefix\url{https://www.sciencedirect.com/science/article/pii/0304885387907219}.

\bibitem[{\citenamefont{Lichtenstein and Katsnelson}(2000)}]{Lichtenstein_PhysRevB_2000}
\bibinfo{author}{\bibfnamefont{A.~I.} \bibnamefont{Lichtenstein}} \bibnamefont{and} \bibinfo{author}{\bibfnamefont{M.~I.} \bibnamefont{Katsnelson}}, \bibinfo{journal}{Phys. Rev. B} \textbf{\bibinfo{volume}{62}}, \bibinfo{pages}{R9283} (\bibinfo{year}{2000}), \urlprefix\url{https://link.aps.org/doi/10.1103/PhysRevB.62.R9283}.

\bibitem[{\citenamefont{Kvashnin et~al.}(2015)\citenamefont{Kvashnin, Gr\aa{}n\"as, Di~Marco, Katsnelson, Lichtenstein, and Eriksson}}]{Igor_PhysRevB_2015}
\bibinfo{author}{\bibfnamefont{Y.~O.} \bibnamefont{Kvashnin}}, \bibinfo{author}{\bibfnamefont{O.}~\bibnamefont{Gr\aa{}n\"as}}, \bibinfo{author}{\bibfnamefont{I.}~\bibnamefont{Di~Marco}}, \bibinfo{author}{\bibfnamefont{M.~I.} \bibnamefont{Katsnelson}}, \bibinfo{author}{\bibfnamefont{A.~I.} \bibnamefont{Lichtenstein}}, \bibnamefont{and} \bibinfo{author}{\bibfnamefont{O.}~\bibnamefont{Eriksson}}, \bibinfo{journal}{Phys. Rev. B} \textbf{\bibinfo{volume}{91}}, \bibinfo{pages}{125133} (\bibinfo{year}{2015}), \urlprefix\url{https://link.aps.org/doi/10.1103/PhysRevB.91.125133}.

\bibitem[{\citenamefont{Kvashnin et~al.}(2020)\citenamefont{Kvashnin, Bergman, Lichtenstein, and Katsnelson}}]{KvashninPhysRevB_2020}
\bibinfo{author}{\bibfnamefont{Y.~O.} \bibnamefont{Kvashnin}}, \bibinfo{author}{\bibfnamefont{A.}~\bibnamefont{Bergman}}, \bibinfo{author}{\bibfnamefont{A.~I.} \bibnamefont{Lichtenstein}}, \bibnamefont{and} \bibinfo{author}{\bibfnamefont{M.~I.} \bibnamefont{Katsnelson}}, \bibinfo{journal}{Phys. Rev. B} \textbf{\bibinfo{volume}{102}}, \bibinfo{pages}{115162} (\bibinfo{year}{2020}), \urlprefix\url{https://link.aps.org/doi/10.1103/PhysRevB.102.115162}.

\bibitem[{\citenamefont{Moriya}(1960)}]{Moriya_PhysRev_1960}
\bibinfo{author}{\bibfnamefont{T.}~\bibnamefont{Moriya}}, \bibinfo{journal}{Phys. Rev.} \textbf{\bibinfo{volume}{120}}, \bibinfo{pages}{91} (\bibinfo{year}{1960}), \urlprefix\url{https://link.aps.org/doi/10.1103/PhysRev.120.91}.

\bibitem[{\citenamefont{Dzyaloshinsky}(1958)}]{DZYALOSHINSKY1958241}
\bibinfo{author}{\bibfnamefont{I.}~\bibnamefont{Dzyaloshinsky}}, \bibinfo{journal}{Journal of Physics and Chemistry of Solids} \textbf{\bibinfo{volume}{4}}, \bibinfo{pages}{241} (\bibinfo{year}{1958}), ISSN \bibinfo{issn}{0022-3697}, \urlprefix\url{https://www.sciencedirect.com/science/article/pii/0022369758900763}.

\bibitem[{\citenamefont{P{\'e}rez~Vicente et~al.}(1999)\citenamefont{P{\'e}rez~Vicente, Tirado, Adouby, Jumas, Tour{\'e}, and Kra}}]{perez_ing_chem.38.9_1999}
\bibinfo{author}{\bibfnamefont{C.}~\bibnamefont{P{\'e}rez~Vicente}}, \bibinfo{author}{\bibfnamefont{J.}~\bibnamefont{Tirado}}, \bibinfo{author}{\bibfnamefont{K.}~\bibnamefont{Adouby}}, \bibinfo{author}{\bibfnamefont{J.}~\bibnamefont{Jumas}}, \bibinfo{author}{\bibfnamefont{A.~A.} \bibnamefont{Tour{\'e}}}, \bibnamefont{and} \bibinfo{author}{\bibfnamefont{G.}~\bibnamefont{Kra}}, \bibinfo{journal}{Inorganic chemistry} \textbf{\bibinfo{volume}{38}}, \bibinfo{pages}{2131} (\bibinfo{year}{1999}).

\bibitem[{\citenamefont{Nakajima}(1963)}]{nakajima_jpcs.24.3_1963}
\bibinfo{author}{\bibfnamefont{S.}~\bibnamefont{Nakajima}}, \bibinfo{journal}{Journal of Physics and Chemistry of Solids} \textbf{\bibinfo{volume}{24}}, \bibinfo{pages}{479} (\bibinfo{year}{1963}), ISSN \bibinfo{issn}{0022-3697}, \urlprefix\url{https://www.sciencedirect.com/science/article/pii/0022369763902075}.

\bibitem[{\citenamefont{Horák et~al.}(1990)\citenamefont{Horák, Stary, Lošťák, and Pancíř}}]{horak_jpcs.51.12_1990}
\bibinfo{author}{\bibfnamefont{J.}~\bibnamefont{Horák}}, \bibinfo{author}{\bibfnamefont{Z.}~\bibnamefont{Stary}}, \bibinfo{author}{\bibfnamefont{P.}~\bibnamefont{Lošťák}}, \bibnamefont{and} \bibinfo{author}{\bibfnamefont{J.}~\bibnamefont{Pancíř}}, \bibinfo{journal}{Journal of Physics and Chemistry of Solids} \textbf{\bibinfo{volume}{51}}, \bibinfo{pages}{1353} (\bibinfo{year}{1990}), ISSN \bibinfo{issn}{0022-3697}, \urlprefix\url{https://www.sciencedirect.com/science/article/pii/002236979090017A}.

\bibitem[{\citenamefont{Zhang et~al.}(2013)\citenamefont{Zhang, Chang, Tang, Zhang, Feng, Li, li~Wang, Chen, Liu, Duan et~al.}}]{zhang_science.339_2013}
\bibinfo{author}{\bibfnamefont{J.}~\bibnamefont{Zhang}}, \bibinfo{author}{\bibfnamefont{C.-Z.} \bibnamefont{Chang}}, \bibinfo{author}{\bibfnamefont{P.}~\bibnamefont{Tang}}, \bibinfo{author}{\bibfnamefont{Z.}~\bibnamefont{Zhang}}, \bibinfo{author}{\bibfnamefont{X.}~\bibnamefont{Feng}}, \bibinfo{author}{\bibfnamefont{K.}~\bibnamefont{Li}}, \bibinfo{author}{\bibfnamefont{L.}~\bibnamefont{li~Wang}}, \bibinfo{author}{\bibfnamefont{X.}~\bibnamefont{Chen}}, \bibinfo{author}{\bibfnamefont{C.}~\bibnamefont{Liu}}, \bibinfo{author}{\bibfnamefont{W.}~\bibnamefont{Duan}}, \bibnamefont{et~al.}, \bibinfo{journal}{Science} \textbf{\bibinfo{volume}{339}}, \bibinfo{pages}{1582} (\bibinfo{year}{2013}), \eprint{https://www.science.org/doi/pdf/10.1126/science.1230905}, \urlprefix\url{https://www.science.org/doi/abs/10.1126/science.1230905}.

\bibitem[{\citenamefont{Zhao et~al.}(2020)\citenamefont{Zhao, Zhang, Mei, Zhou, Yi, Zhang, Yu, Xiao, Wang, Samarth et~al.}}]{zhao_nature.588_2020}
\bibinfo{author}{\bibfnamefont{Y.-F.} \bibnamefont{Zhao}}, \bibinfo{author}{\bibfnamefont{R.}~\bibnamefont{Zhang}}, \bibinfo{author}{\bibfnamefont{R.}~\bibnamefont{Mei}}, \bibinfo{author}{\bibfnamefont{L.-J.} \bibnamefont{Zhou}}, \bibinfo{author}{\bibfnamefont{H.}~\bibnamefont{Yi}}, \bibinfo{author}{\bibfnamefont{Y.-Q.} \bibnamefont{Zhang}}, \bibinfo{author}{\bibfnamefont{J.}~\bibnamefont{Yu}}, \bibinfo{author}{\bibfnamefont{R.}~\bibnamefont{Xiao}}, \bibinfo{author}{\bibfnamefont{K.}~\bibnamefont{Wang}}, \bibinfo{author}{\bibfnamefont{N.}~\bibnamefont{Samarth}}, \bibnamefont{et~al.}, \bibinfo{journal}{Nature} \textbf{\bibinfo{volume}{588}}, \bibinfo{pages}{419} (\bibinfo{year}{2020}).

\bibitem[{\citenamefont{Wang and Chang}(2022)}]{Wang_2022-GKA}
\bibinfo{author}{\bibfnamefont{M.-C.} \bibnamefont{Wang}} \bibnamefont{and} \bibinfo{author}{\bibfnamefont{C.-R.} \bibnamefont{Chang}}, \bibinfo{journal}{Journal of The Electrochemical Society} \textbf{\bibinfo{volume}{169}}, \bibinfo{pages}{053507} (\bibinfo{year}{2022}), \urlprefix\url{https://dx.doi.org/10.1149/1945-7111/ac7006}.

\bibitem[{\citenamefont{Jungwirth et~al.}(2006)\citenamefont{Jungwirth, Sinova, Ma\ifmmode~\check{s}\else \v{s}\fi{}ek, Ku\ifmmode~\check{c}\else \v{c}\fi{}era, and MacDonald}}]{TJungwirth_RevModPhys.78.809_2006}
\bibinfo{author}{\bibfnamefont{T.}~\bibnamefont{Jungwirth}}, \bibinfo{author}{\bibfnamefont{J.}~\bibnamefont{Sinova}}, \bibinfo{author}{\bibfnamefont{J.}~\bibnamefont{Ma\ifmmode~\check{s}\else \v{s}\fi{}ek}}, \bibinfo{author}{\bibfnamefont{J.}~\bibnamefont{Ku\ifmmode~\check{c}\else \v{c}\fi{}era}}, \bibnamefont{and} \bibinfo{author}{\bibfnamefont{A.~H.} \bibnamefont{MacDonald}}, \bibinfo{journal}{Rev. Mod. Phys.} \textbf{\bibinfo{volume}{78}}, \bibinfo{pages}{809} (\bibinfo{year}{2006}), \urlprefix\url{https://link.aps.org/doi/10.1103/RevModPhys.78.809}.

\bibitem[{\citenamefont{Sato et~al.}(2010)\citenamefont{Sato, Bergqvist, Kudrnovsk\'y, Dederichs, Eriksson, Turek, Sanyal, Bouzerar, Katayama-Yoshida, Dinh et~al.}}]{KSato_RevModPhys.82.1633_2010}
\bibinfo{author}{\bibfnamefont{K.}~\bibnamefont{Sato}}, \bibinfo{author}{\bibfnamefont{L.}~\bibnamefont{Bergqvist}}, \bibinfo{author}{\bibfnamefont{J.}~\bibnamefont{Kudrnovsk\'y}}, \bibinfo{author}{\bibfnamefont{P.~H.} \bibnamefont{Dederichs}}, \bibinfo{author}{\bibfnamefont{O.}~\bibnamefont{Eriksson}}, \bibinfo{author}{\bibfnamefont{I.}~\bibnamefont{Turek}}, \bibinfo{author}{\bibfnamefont{B.}~\bibnamefont{Sanyal}}, \bibinfo{author}{\bibfnamefont{G.}~\bibnamefont{Bouzerar}}, \bibinfo{author}{\bibfnamefont{H.}~\bibnamefont{Katayama-Yoshida}}, \bibinfo{author}{\bibfnamefont{V.~A.} \bibnamefont{Dinh}}, \bibnamefont{et~al.}, \bibinfo{journal}{Rev. Mod. Phys.} \textbf{\bibinfo{volume}{82}}, \bibinfo{pages}{1633} (\bibinfo{year}{2010}), \urlprefix\url{https://link.aps.org/doi/10.1103/RevModPhys.82.1633}.

\bibitem[{\citenamefont{Wang et~al.}(2016)\citenamefont{Wang, Fan, Zhu, and Wu}}]{HWang_EPL.114_2016}
\bibinfo{author}{\bibfnamefont{H.}~\bibnamefont{Wang}}, \bibinfo{author}{\bibfnamefont{F.}~\bibnamefont{Fan}}, \bibinfo{author}{\bibfnamefont{S.}~\bibnamefont{Zhu}}, \bibnamefont{and} \bibinfo{author}{\bibfnamefont{H.}~\bibnamefont{Wu}}, \bibinfo{journal}{Europhysics Letters} \textbf{\bibinfo{volume}{114}}, \bibinfo{pages}{47001} (\bibinfo{year}{2016}), \urlprefix\url{https://dx.doi.org/10.1209/0295-5075/114/47001}.

\bibitem[{\citenamefont{Xue et~al.}(2019)\citenamefont{Xue, Hou, Wang, and Wu}}]{XFeng_PRB.100_2019}
\bibinfo{author}{\bibfnamefont{F.}~\bibnamefont{Xue}}, \bibinfo{author}{\bibfnamefont{Y.}~\bibnamefont{Hou}}, \bibinfo{author}{\bibfnamefont{Z.}~\bibnamefont{Wang}}, \bibnamefont{and} \bibinfo{author}{\bibfnamefont{R.}~\bibnamefont{Wu}}, \bibinfo{journal}{Phys. Rev. B} \textbf{\bibinfo{volume}{100}}, \bibinfo{pages}{224429} (\bibinfo{year}{2019}), \urlprefix\url{https://link.aps.org/doi/10.1103/PhysRevB.100.224429}.

\bibitem[{\citenamefont{Peixoto et~al.}(2020{\natexlab{b}})\citenamefont{Peixoto, Bentmann, Rüßmann, Tcakaev, Winnerlein, Schreyeck, Schatz, Vidal, Stier, Zabolotnyy et~al.}}]{TRFPeixoto_NJPQM.5_2020}
\bibinfo{author}{\bibfnamefont{T.~R.~F.} \bibnamefont{Peixoto}}, \bibinfo{author}{\bibfnamefont{H.}~\bibnamefont{Bentmann}}, \bibinfo{author}{\bibfnamefont{P.}~\bibnamefont{Rüßmann}}, \bibinfo{author}{\bibfnamefont{A.-V.} \bibnamefont{Tcakaev}}, \bibinfo{author}{\bibfnamefont{M.}~\bibnamefont{Winnerlein}}, \bibinfo{author}{\bibfnamefont{S.}~\bibnamefont{Schreyeck}}, \bibinfo{author}{\bibfnamefont{S.}~\bibnamefont{Schatz}}, \bibinfo{author}{\bibfnamefont{R.~C.} \bibnamefont{Vidal}}, \bibinfo{author}{\bibfnamefont{F.}~\bibnamefont{Stier}}, \bibinfo{author}{\bibfnamefont{V.}~\bibnamefont{Zabolotnyy}}, \bibnamefont{et~al.}, \bibinfo{journal}{npj quantum materials} \textbf{\bibinfo{volume}{5}}, \bibinfo{pages}{87} (\bibinfo{year}{2020}{\natexlab{b}}), ISSN \bibinfo{issn}{2397-4648}, \urlprefix\url{https://juser.fz-juelich.de/record/888146}.

\bibitem[{\citenamefont{Choi et~al.}(2012)\citenamefont{Choi, Jo, Lee, Lee, Jo, Kajino, Takabatake, Ko, Park, and Jung}}]{Ychoi_APL.101_2012}
\bibinfo{author}{\bibfnamefont{Y.~H.} \bibnamefont{Choi}}, \bibinfo{author}{\bibfnamefont{N.~H.} \bibnamefont{Jo}}, \bibinfo{author}{\bibfnamefont{K.~J.} \bibnamefont{Lee}}, \bibinfo{author}{\bibfnamefont{H.~W.} \bibnamefont{Lee}}, \bibinfo{author}{\bibfnamefont{Y.~H.} \bibnamefont{Jo}}, \bibinfo{author}{\bibfnamefont{J.}~\bibnamefont{Kajino}}, \bibinfo{author}{\bibfnamefont{T.}~\bibnamefont{Takabatake}}, \bibinfo{author}{\bibfnamefont{K.-T.} \bibnamefont{Ko}}, \bibinfo{author}{\bibfnamefont{J.-H.} \bibnamefont{Park}}, \bibnamefont{and} \bibinfo{author}{\bibfnamefont{M.~H.} \bibnamefont{Jung}}, \bibinfo{journal}{Applied Physics Letters} \textbf{\bibinfo{volume}{101}}, \bibinfo{pages}{152103} (\bibinfo{year}{2012}).

\bibitem[{\citenamefont{Li et~al.}(2014)\citenamefont{Li, Zou, Li, and Zhou}}]{Yli_JCP.140_2014}
\bibinfo{author}{\bibfnamefont{Y.}~\bibnamefont{Li}}, \bibinfo{author}{\bibfnamefont{X.}~\bibnamefont{Zou}}, \bibinfo{author}{\bibfnamefont{J.}~\bibnamefont{Li}}, \bibnamefont{and} \bibinfo{author}{\bibfnamefont{G.}~\bibnamefont{Zhou}}, \bibinfo{journal}{The Journal of Chemical Physics} \textbf{\bibinfo{volume}{140}}, \bibinfo{pages}{124704} (\bibinfo{year}{2014}).

\bibitem[{\citenamefont{Almeleh and Goldstein}(1962)}]{Nalmeleh_PRB.128_1962}
\bibinfo{author}{\bibfnamefont{N.}~\bibnamefont{Almeleh}} \bibnamefont{and} \bibinfo{author}{\bibfnamefont{B.}~\bibnamefont{Goldstein}}, \bibinfo{journal}{Phys. Rev.} \textbf{\bibinfo{volume}{128}}, \bibinfo{pages}{1568} (\bibinfo{year}{1962}), \urlprefix\url{https://link.aps.org/doi/10.1103/PhysRev.128.1568}.

\bibitem[{\citenamefont{Szczytko et~al.}(1999)\citenamefont{Szczytko, Twardowski, \ifmmode \acute{S}\else \'{S}\fi{}wia\ifmmode~\mbox{\c{}}\else \c{}\fi{}tek, Palczewska, Tanaka, Hayashi, and Ando}}]{JSzczytko_PPRB.60_1999}
\bibinfo{author}{\bibfnamefont{J.}~\bibnamefont{Szczytko}}, \bibinfo{author}{\bibfnamefont{A.}~\bibnamefont{Twardowski}}, \bibinfo{author}{\bibfnamefont{K.}~\bibnamefont{\ifmmode \acute{S}\else \'{S}\fi{}wia\ifmmode~\mbox{\c{}}\else \c{}\fi{}tek}}, \bibinfo{author}{\bibfnamefont{M.}~\bibnamefont{Palczewska}}, \bibinfo{author}{\bibfnamefont{M.}~\bibnamefont{Tanaka}}, \bibinfo{author}{\bibfnamefont{T.}~\bibnamefont{Hayashi}}, \bibnamefont{and} \bibinfo{author}{\bibfnamefont{K.}~\bibnamefont{Ando}}, \bibinfo{journal}{Phys. Rev. B} \textbf{\bibinfo{volume}{60}}, \bibinfo{pages}{8304} (\bibinfo{year}{1999}), \urlprefix\url{https://link.aps.org/doi/10.1103/PhysRevB.60.8304}.

\bibitem[{\citenamefont{Sasaki et~al.}(2002)\citenamefont{Sasaki, Liu, Furdyna, Palczewska, Szczytko, and Twardowski}}]{YSasaki_JAP.91_2022}
\bibinfo{author}{\bibfnamefont{Y.}~\bibnamefont{Sasaki}}, \bibinfo{author}{\bibfnamefont{X.}~\bibnamefont{Liu}}, \bibinfo{author}{\bibfnamefont{J.~K.} \bibnamefont{Furdyna}}, \bibinfo{author}{\bibfnamefont{M.}~\bibnamefont{Palczewska}}, \bibinfo{author}{\bibfnamefont{J.}~\bibnamefont{Szczytko}}, \bibnamefont{and} \bibinfo{author}{\bibfnamefont{A.}~\bibnamefont{Twardowski}}, \bibinfo{journal}{Journal of Applied Physics} \textbf{\bibinfo{volume}{91}}, \bibinfo{pages}{7484} (\bibinfo{year}{2002}), ISSN \bibinfo{issn}{0021-8979}, \eprint{https://pubs.aip.org/aip/jap/article-pdf/91/10/7484/8060634/7484\_1\_online.pdf}, \urlprefix\url{https://doi.org/10.1063/1.1447214}.

\bibitem[{\citenamefont{Kvashnin et~al.}(2016)\citenamefont{Kvashnin, Cardias, Szilva, Di~Marco, Katsnelson, Lichtenstein, Nordstr\"om, Klautau, and Eriksson}}]{YOKvashnin_PRL_116.2016}
\bibinfo{author}{\bibfnamefont{Y.~O.} \bibnamefont{Kvashnin}}, \bibinfo{author}{\bibfnamefont{R.}~\bibnamefont{Cardias}}, \bibinfo{author}{\bibfnamefont{A.}~\bibnamefont{Szilva}}, \bibinfo{author}{\bibfnamefont{I.}~\bibnamefont{Di~Marco}}, \bibinfo{author}{\bibfnamefont{M.~I.} \bibnamefont{Katsnelson}}, \bibinfo{author}{\bibfnamefont{A.~I.} \bibnamefont{Lichtenstein}}, \bibinfo{author}{\bibfnamefont{L.}~\bibnamefont{Nordstr\"om}}, \bibinfo{author}{\bibfnamefont{A.~B.} \bibnamefont{Klautau}}, \bibnamefont{and} \bibinfo{author}{\bibfnamefont{O.}~\bibnamefont{Eriksson}}, \bibinfo{journal}{Phys. Rev. Lett.} \textbf{\bibinfo{volume}{116}}, \bibinfo{pages}{217202} (\bibinfo{year}{2016}), \urlprefix\url{https://link.aps.org/doi/10.1103/PhysRevLett.116.217202}.

\end{thebibliography}

\end{document}



\title{Supplementary material for the manuscript entitled\\``Revisiting magnetic exchange interactions in transition metal
doped Bi$_2$Se$_3$ using DFT+MFT"}
\author{Sagar Sarkar}
\affiliation{Asia Pacific Center for Theoretical Physics, Pohang, 37673, Korea}
\affiliation{Department of Physics and Astronomy, Uppsala University, Uppsala, 751 20, Sweden}
\author{Shivalika Sharma}
\affiliation{Asia Pacific Center for Theoretical Physics, Pohang, 37673, Korea}
\affiliation{Institute of Physics, Nicolaus Copernicus University, 87-100 Toruń, Poland}

\author{Igor Di Marco}
\affiliation{Asia Pacific Center for Theoretical Physics, Pohang, 37673, Korea}
\affiliation{Institute of Physics, Nicolaus Copernicus University, 87-100 Toruń, Poland}

\date{January 2024}

\maketitle

\section{Effect of local correlation (U), Spin-Orbit coupling (SOC), and U+SOC induced changes on the electronic structure.}

\begin{figure}[h]
\includegraphics[scale=0.5]{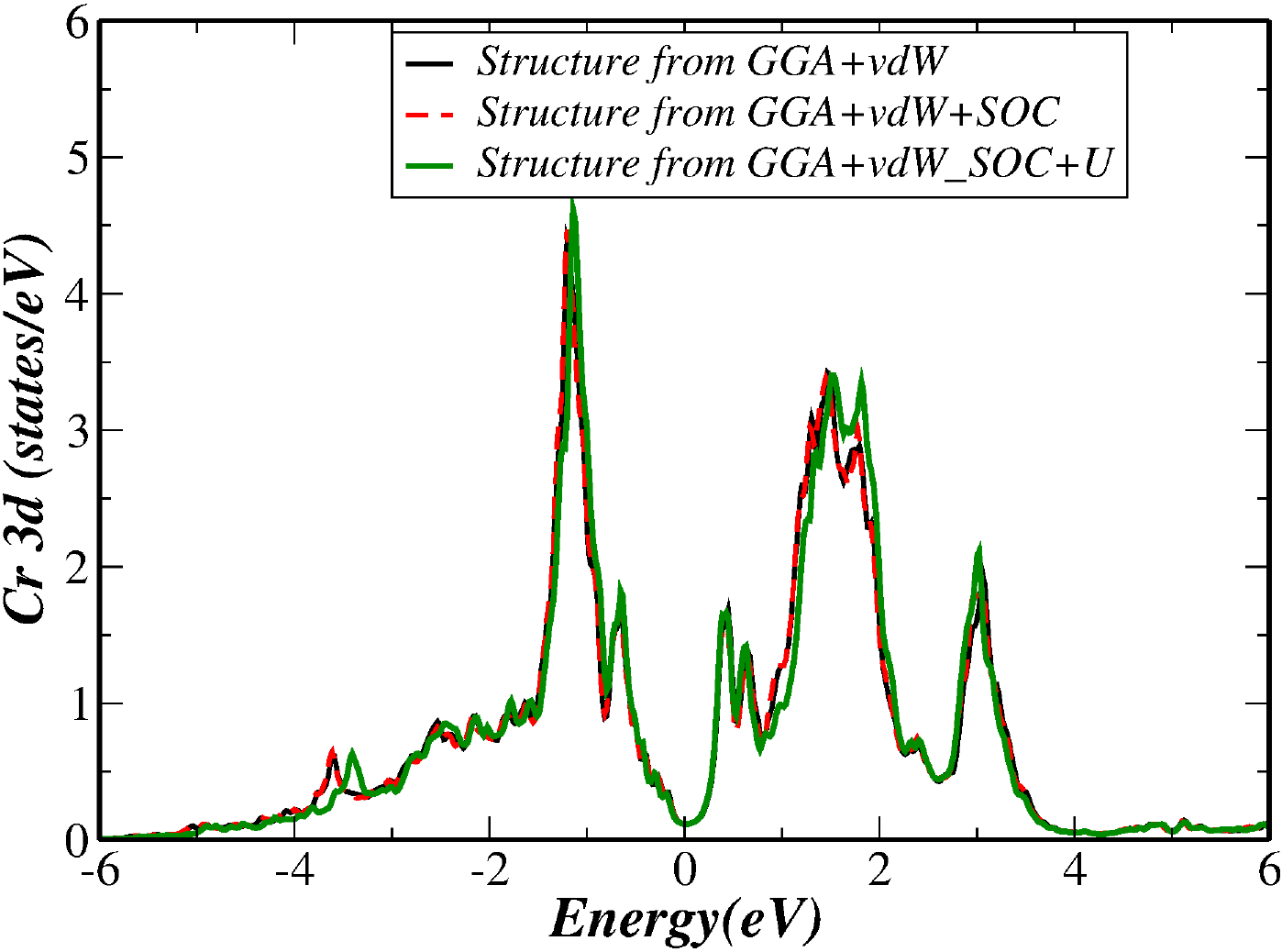}
\caption{Comparison between the local pdos corresponding to the Cr $3d$ states in three different structures. The same calculation settings are used for all the calculations.}
\label{FigS1}
\end{figure}
\FloatBarrier

While doing the structural optimization of our TM doped Bi$_2$Se$_3$ using different computational schemes using VASP, we found that the local correlation (U) and Spin-Orbit Coupling (SOC) induce small local structural changes over the GGA+vdW optimized structures. To check the impact of these additional small changes on the electronic structure, we considered the GGA+vdW, GGA+vdW+SOC, and GGA+vdW+SOC+U optimized structures of the Cr-doped system from VASP calculations. We then perform plain LDA calculation in RSPt with the same computational settings for the three structures. The comparison of the Cr $3d$ pdos from the three calculations is shown in the above figure. We do not see any significant difference between them supporting our claim that the additional small structural changes induced by SOC and U are not so important and could be ignored. The major structural changes as a result of doping are well captured in all the computational schemes including GGA+vdW. Hence, we decided to do the structural optimization of our doped systems using this scheme in VASP to save time and computational resources.

\section{Comparison of the band structure of bulk pristine B\MakeLowercase{i}$_2$S\MakeLowercase{e}$_3$ calculated in RSPt, VASP, and from literature.}

\begin{figure}[h]
\includegraphics[scale=0.55]{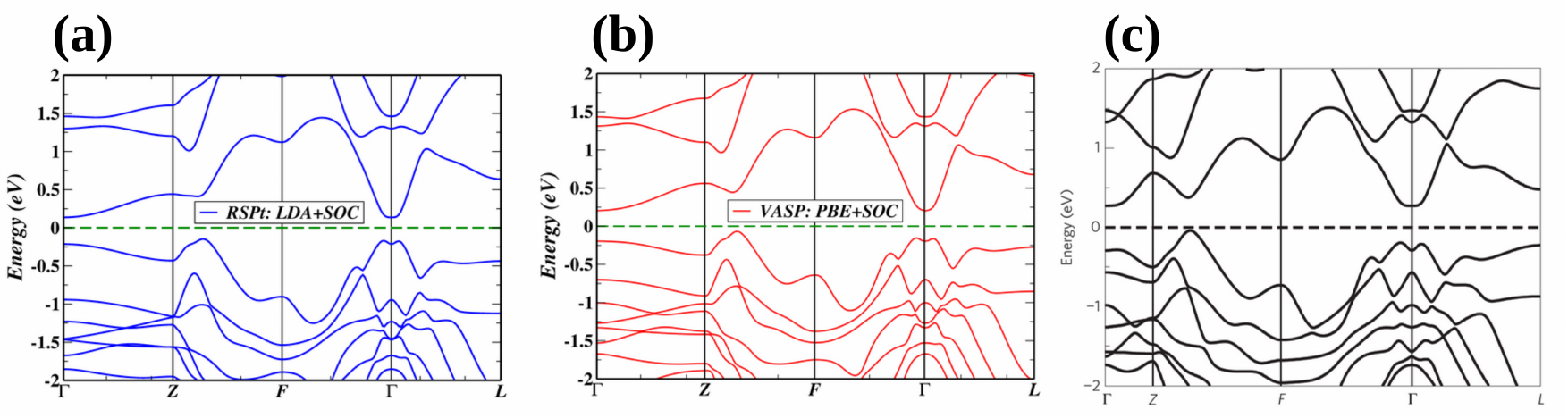}
\caption{Band structure of primitive Rhombohedral Bi$_2$Se$_3$  with SOC in (a) RSPt using LDA Ex-corr functional, (b) VASP using GGA-PBE Ex-corr functional. (c) Band structure of the same from Ref. [1].}
\label{FigS2}
\end{figure}
\FloatBarrier

As we can see from above, the band structure calculated using LDA Ex-corr functional in RSPt agrees very well with the one calculated using GGA-PBE in VASP and as reported in the previous work from Nature Physics.

\section{Structural changes in the 100\% doping case and 50\% doping case in a comparative manner.}

\begin{figure}[h]
\includegraphics[scale=0.55]{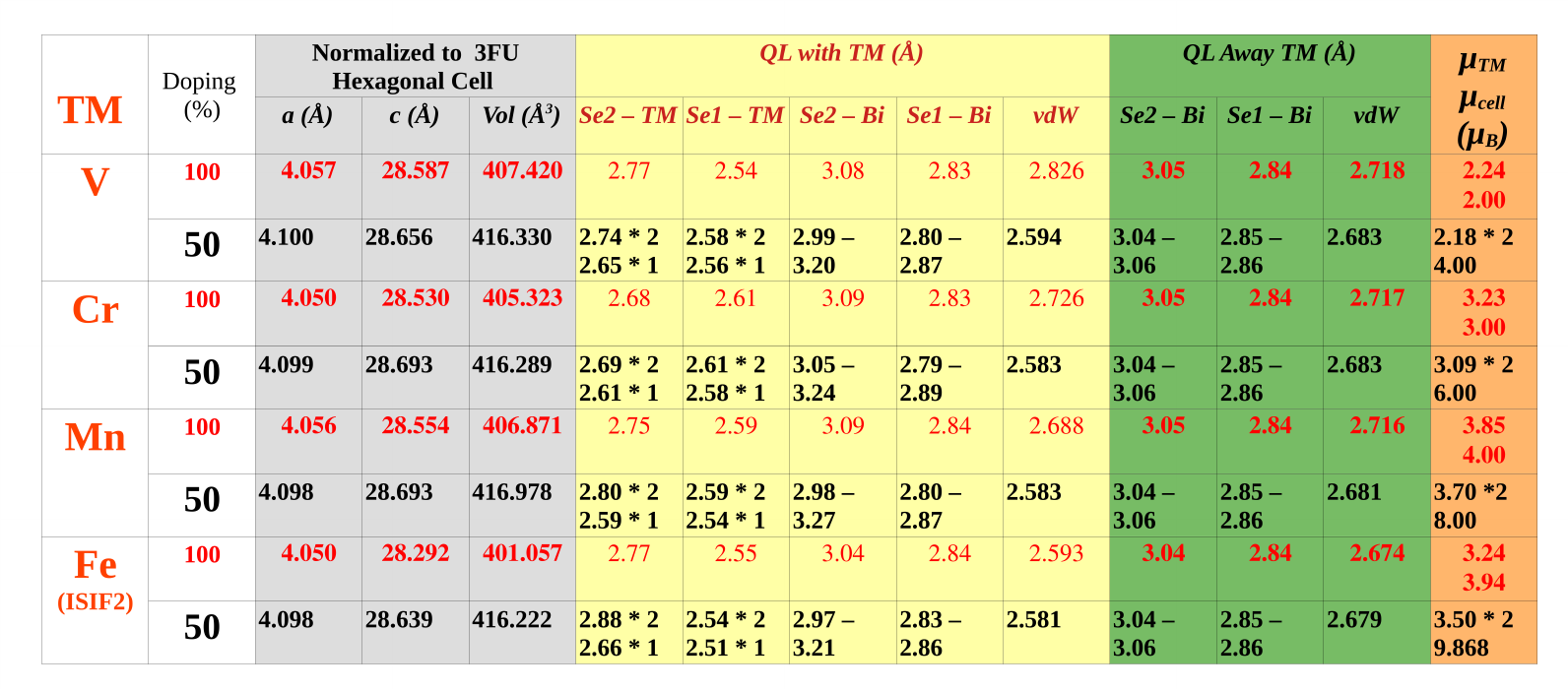}
\caption{A table showing the cation-anion bonds in the TM doped Bi$_2$Se$_3$ series in a comparative manner for 100\% and 50\% doping cases.}
\label{FigS3}
\end{figure}
\FloatBarrier

In the 50\% doping case the lattice parameters and system volume increase as a result of more Bi content. Also in this case, the unit cell volume and lattice parameters come out to be almost the same for each dopant type (V, Cr, Mn, and Fe). This is something different from the 100\% doping case but is also expected because, for lower doping like this, the volumetric properties shall depend more on the host and not on the dopants.  
In the QL with the doped TM atoms, we see 2 larger, and 1 smaller TM-Se bonds in both the Se layer due to local distortions and spacial asymmetry.  Also, there is a disorder in the Bi-Se bonds giving a range of values. The undoped QLs away from TM atoms again remain unaffected as before.

\section{References}

[1] H. Zhang et al, Nature Physics volume 5, pages 438–442 (2009).